\documentclass[12pt]{article}
\usepackage{amsmath}
\usepackage{amsfonts}
\usepackage{amssymb}
\usepackage{graphics}
\usepackage{epsfig}
\usepackage{color,amsmath,bm}
\usepackage{graphicx}
\usepackage{cite}

\newcommand{\beq}{\begin{equation}}
\newcommand{\eeq}{\end{equation}}

\def\beqn{\begin{eqnarray}}
\def\eeqn{\end{eqnarray}}
\def\lsim{\mathrel{\rlap{\lower3pt\hbox{\hskip0pt$\sim$}}
    \raise1pt\hbox{$<$}}}
\def\gsim{\mathrel{\rlap{\lower4pt\hbox{\hskip1pt$\sim$}}
    \raise1pt\hbox{$>$}}}
\def\ntwo{${\mathcal N}=2\,$}

\begin{document}

\begin{titlepage}
\begin{flushright}
FTPI-MINN-07/16\\
UMN-TH-2603/07\\
May 2/2007

\end{flushright}

\vspace{0.5cm}

\begin{center}

{\Large\bf  $Q$ Torus in \ntwo SQED  }

\vspace{1cm}

{\large\bf
S.\ Bolognesi$^{\,\diamond\,\star}$ {\small and} M.\ Shifman$^{\,\star}$  \\[2mm]
{\small \sl $^{\,\diamond}$The Niels Bohr Institute, Blegdamsvej 17,}\\[-1mm]
{\small \sl  DK-2100, Copenhagen \O, Denmark, and}\\[-1mm]
{\small \sl University of Southern Denmark, DK-5230 Odense, Denmark}
\\
{\small \sl $^{\,\star}$William I. Fine Theoretical Physics Institute},\\[-1mm]
{\small \sl University of Minnesota, Minneapolis, MN 55455} }

\end{center}

\vspace{3mm}

\begin{abstract}
We construct ``Flying Saucer" solitons in supersymmetric
\mbox{${\mathcal N}=2$} gauge theory which is known to support BPS
domain walls with a U(1) gauge field localized on its worldvolume.
We demonstrate that this model supports exotic particle-like
solitons with the shape of a torus.
$Q$ tori, and also similar solitons of higher genera, are obtained
by folding the domain wall into an appropriate surface. Nontrivial
cycles on the domain wall worldvolume (handles) are stabilized by
{\it crossed} electric and magnetic fields inside the folded domain
wall. Three distinct frameworks are used to prove the existence of
these ``Flying Saucer" solitons and  study their properties: the
worldvolume description (including the Dirac--Born--Infeld action),
the bulk theory description in the sigma-model limit, and the bulk
theory description in the thin-edge approximation. In the
sigma-model framework the  $Q$ torus is shown to be related to the
Hopf Skyrmion studied previously.

\end{abstract}

\end{titlepage}

\newpage

\tableofcontents

\section{Introduction}

\label{intro}

In this paper we show that the benchmark ${\mathcal{N}}=2$ gauge theory \cite%
{Shifman2002}, which was known to support domain walls with a U(1)
gauge field localized on the wall and the Abrikosov--Nielsen--Olesen
flux tubes, also gives rise to exotic solitons whose topology is
that of a torus (we also suggest the existence of arbitrary genus
solitons). The stability is due to the existence of a number of
conserved quantum numbers. One of them, $q$, is the U(1) charge on
the surface; others $n, N, ...$ describe magnetic fluxes
corresponding to various cycles. For instance, for a toroidal
soliton, there are two cycles and, hence, two magnetic fluxes.

The general idea behind the solitons of the ``Flying Saucer" type is
as follows. We start from a ``plane" domain wall (which is 1/2
BPS-saturated and topologically stable), and then fold it to make
surfaces of arbitrary genera. These surfaces are closed and
orientable. To stabilize all handles of the surfaces we let magnetic
fluxes flow inside the wall. Each noncontractable cycle is
characterized by its own flux. For instance, in the case of the $Q$
torus\footnote{We use a convention a la Coleman
\cite{Coleman:1985ki} and denote Q-solitons exitations that carry a
charge of a  conserved perturbative current.}, two magnetic fluxes
must be nonvanishing. Magnetic fluxes by themselves are insufficient
for complete stabilization. In addition to the magnetic fluxes the
solitons under consideration must be $Q$ charged. The gauge U(1)
symmetry is Higgsed in the bulk; therefore the electric field is
screened in the bulk. However, a global U(1) survives providing for
a possibility of $Q$ charging. Inside the wall we are in the Coulomb
phase; therefore an electric field perpendicular to the wall surface
can develop. It does develop if we endow the ``Flying Saucer"
soliton with the $Q$ charge. The combined action of the magnetic
fluxes and the $Q$ charge leads to full stabilization.

We exploit three distinct (although not completely unrelated) frameworks to
carry out our analysis: the worldvolume description (including that based on
the Dirac--Born--Infeld action), the bulk theory description in the
thin-edge approximation and the bulk theory description in the sigma-model
limit. All three considerations provide a proof of the existence of the $Q$
solitons and allow us to study some properties of the $Q$ torus --- the $Q$
soliton on which we mostly focus in the present paper. In the sigma-model
framework it becomes clear that the $Q$ torus is in fact related to the Hopf
Skyrmion studied previously.

Our basic bulk theory is ${\mathcal{N}}=2$ supersymmetric $U(1)$ gauge
theory with two ``quark" hypermultiplets $Q^{A}$, $A=1,2$ and the
Fayet--Iliopoulos term. Each hypermultiplet consists of two ${\mathcal{N}}=1$
chiral superfields, $Q^A$ and $\tilde Q_A$ ($A=1,2)$, with the electric
charges $+1/2$ and $-1/2$, respectively. The quark mass terms are introduced
in the superpotential as follows
\begin{equation}
{\mathcal{W}}_m = \sum_{A=1,2}\, m_A \, Q^A\, \tilde Q_A\,.
\end{equation}
The existence of walls requires $m_1\neq m_2$. We will assume $m_{1,2}$ to
be real and positive, and
\begin{equation}
\Delta m\equiv |m_1-m_2| \ll m_{1,2}\,.
\end{equation}
Moreover, we will introduce the Fayet--Iliopoulos term. In Ref.~\cite%
{Shifman2002} it was introduced through a superpotential; following \cite%
{Sakai2005} we will introduce it here through a $D$ term. These two methods
reduce to each other identically \cite{Hanany,Vainshtein}. An ${\mathcal{N}}%
=1$ chiral superfield ${\mathcal{A}}$ (with the lowest component $a$) ---
the ${\mathcal{N}}=2$ superpartner of the gauge field strength tensor $W$
--- is also a part of our bulk theory.


Organization of the paper is as follows. In Sect.~\ref{bulkth} we
briefly review relevant aspects of the bulk theory. In Sect.
\ref{fieldexited} we consider the domain wall with fields ewxited on
on its world volume. In ~\ref{cmfitw} we explain how the magnetic
flux can propagate inside the wall. (In the dual Polyakov language
this is an electric field on the wall). In ~\ref{mean} we introduce
a $Q$ charging procedure. We explain that the wall can be endowed
with a conserved (global) U(1) charge which generates an electric
field inside the wall perpendicular to the wall surface. (In the
dual Polyakov language this is a magnetic field on the wall).
Section~\ref{cylinder} presents a $Q$ cylinder
--- a $Q$ wall folded in the form of a cylinder.
Subsect.~\ref{qcbstp} treats the $Q$ cylinder from the bulk-theory
standpoint, while in ~\ref{implqc} the same object is analyzed in
the sigma-model framework. These two complementary considerations
are confronted in Subsect.~\ref{cbtsma}. A new crucial element on
the road to the $Q$ torus
--- twisting --- is introduced in Subsect.~\ref{ttc}. A concept design of the $%
Q $ torus is presented in Sect.~\ref{tqct}. We discuss the $Q$ torus
in the framework of the bulk theory in Subsect.~\ref{qtibt}. Then in
~\ref{qtsml} we continue this discussion in the sigma-model limit.
Subsection~\ref{comone} is a comment on stability of the $Q$ torus
while Subsect.~\ref{isahg} addresses the issue of stability of the
$Q$ solitons of higher genera. This discussion is continued in
~\ref{hige}. Subsect.~\ref{qchar} is devoted to stability of the $Q$
torus with respect to leakage of the $Q$ charge in the
bulk through radiation of elementary quanta. In Sect.~\ref{tdbia}  we use the Dirac--Born--Infeld action to further explore the $Q$%
-toric solitons in the framework of the worldvolume theory. Finally, Sect.~%
\ref{conoq} summarizes our conclusions and lists some open questions.


\section{The bulk theory: a brief review of known facts}

\label{bulkth}

The simplest model suitable for our construction is ${\mathcal{N}} =2$ SQED
with two matter flavors. It was first used in the analysis of BPS domain
walls supporting gauge fields on the wall surface in Ref.~\cite{Shifman2002}%
. We refer the reader to \cite{Shifman2002} and the review paper \cite{rp}
for a detailed description of the model. Here we will just outline basic
features which will be necessary in what follows.

${\mathcal{N}}=2$ supersymmetric U(1) gauge theory under consideration has
two matter hypermultiplets $Q^{A}$, $A=1,2$. Each hypermultiplet consists of
two ${\mathcal{N}}=1$ chiral superfields. We will introduce the following ${%
\mathcal{N}}=1$ chiral superfields:
\begin{equation*}
Q^A \,\,\, \mbox{ and} \,\,\, \tilde Q_A
\end{equation*}
($A=1,2)$, with the electric charges $+1/2$ and $-1/2$, respectively. The
quark mass terms are introduced via the superpotential,
\begin{equation}
{\mathcal{W}}_m = \sum_{A=1,2}\, m_A \, Q^A\, \tilde Q_A\,.
\end{equation}

The bosonic part of the action of this theory is
\begin{eqnarray}  \label{n2sqed}
&& S=\int d^4 x \left\{ -\frac{1}{4 g^2} F_{\mu \nu}^2 + \frac{1}{g^2}
|\partial_\mu a|^2 +\bar{\nabla}_\mu \bar{q}_A \nabla_\mu q^A + \bar{\nabla}%
_\mu \tilde{q}_A \nabla_\mu \bar{\tilde{q}}^A \right.  \notag \\[3mm]
&& \left.-\left[ \frac{g^2}{8}\left(|q^A|^2-|\tilde{q}_A|^2-\xi\right)^2 +
\frac{g^2}{2} \left|\tilde{q}_A q^A\right|^2 +\frac{1}{2}(|q^A|^2+ |\tilde{q}
^A|^2)\left| a+\sqrt{2}m_A\right|^2 \right] \right\},  \notag \\
\end{eqnarray}
where
\begin{equation}
\nabla_\mu=\partial_\mu-\frac{i}{2}A_\mu\,,\qquad \bar{\nabla}
_\mu=\partial_\mu+\frac{i}{2}A_\mu\,,  \label{nab}
\end{equation}
and $q,\,\,\tilde q$ are the scalar fields from $Q,\,\,\tilde Q$. Moreover, $%
\xi$ is the coefficient in front of the Fayet--Iliopoulos (FI) $D$ term, and
$g$ is the U(1) gauge coupling. This is the simplest case which admits
domain wall interpolating between two quark vacua.

There are two vacua in this theory: in the first vacuum
\begin{equation}
a=-\sqrt{2}\, m_1,\quad q^1=\sqrt{\xi}, \quad q^2=0\,,\quad \tilde{q}=0\,,
\label{fv}
\end{equation}
and in the second one
\begin{equation}
a=-\sqrt{2} \, m_2, \quad q^1=0, \quad q^2=\sqrt{\xi}, \quad \tilde{q}=0\,.
\label{sv}
\end{equation}
The vacuum expectation value (VEV) of the field $\tilde{q}$ vanishes in both
vacua, so we can disregard $\tilde{q}$ at the classical level.


Let us discuss the mass spectrum of light fields in both quark vacua.
Consider for definiteness the first vacuum, Eq.~(\ref{fv}). The spectrum can
be obtained by diagonalizing the quadratic form in the above Lagrangian~\cite%
{Shifman2002}. The result is as follows: one real component of the field $%
q^1 $ is eaten up by the Higgs mechanism to become the third components of
the massive photon. Three components of the massive photon, one remaining
component of $q^1$, and four real components of the fields $\tilde{q}_1$ and
$a$ form one long ${\mathcal{N}}=2\,$ multiplet (8 boson states + 8 fermion
states), with mass\,\footnote{%
The constraint (\ref{mggxi}) is equivalent to $\Delta m \gg m_\gamma$.}
\begin{equation}  \label{mgammaw}
m_{\gamma}^2=\frac12\, g^2\,\xi.
\end{equation}
The second flavor $q^2$, $\tilde{q}_2$ (which does not condense in this
vacuum) forms one short ${\mathcal{N}}=2\,$ multiplet (4 boson states + 4
fermion states), with mass $\Delta m$ which is heavier than the mass of the
vector supermultiplet. In the second vacuum the mass spectrum is similar ---
the roles of the first and the second flavors are interchanged. The BPS
domain wall which is a starting point of our construction interpolates
between these two vacua. The wall solution is discussed in detail in~\cite%
{Shifman2002}. We reiterate here qualitative features, see Fig.~\ref%
{syfigthree}.


The mass parameters $m_1,m_2$ are assumed to be real. One can consider
various limits for the ratio $|\Delta m|/\sqrt\xi $. First, we will assume
\begin{equation}
| \Delta m | \gg g\sqrt{\xi} \,.  \label{mggxi}
\end{equation}
In this approximation the domain wall has a well-defined \textit{three-layer
structure}: two edges whose thickness $\delta$ is small, $\delta\sim
m_\gamma^{-1}$, and a wide middle domain whose thickness $d$ is large, $%
d\sim (\Delta m/m_\gamma )\,m_\gamma^{-1}$, see below. The thickness $d$, as
well as the wall tension can be obtained as follows. First of all, one can
neglect the edges $E_{1,2}$. This is a valid approximation, to be referred
to as the thin-edge approximation. There are two contributions to the wall
energy from the middle domain $M$: the kinetic energy of the $a$ field $%
|\Delta\, a |^2/(g^2\,d)$ and the loss of energy in the Coulomb vacuum $%
q=\tilde q =0$ in the middle of the wall, $g^2\,\xi^2\, d/8$. We must
minimize the expression for the wall tension
\begin{equation}
T_{\mathrm{w}} = \frac{2 (\Delta m)^2}{g^2\,d} +\frac{g^2\,\xi^2\, d}{8}
\label{2terms}
\end{equation}
with respect to $d$, which yields
\begin{equation}
d = \frac{4\Delta m}{g^2\xi}= \frac{2\Delta m}{m_\gamma^2}\,,  \label{d}
\end{equation}
and
\begin{equation}
T_{\mathrm{w}}=\xi \, \Delta m\,.  \label{wten}
\end{equation}
Although at first sight Eq.~(\ref{wten}) might seem approximate, in fact it
is exact \cite{Shifman2002}.
\begin{figure}[h!tb]
\epsfxsize=7cm \centerline{\epsfbox{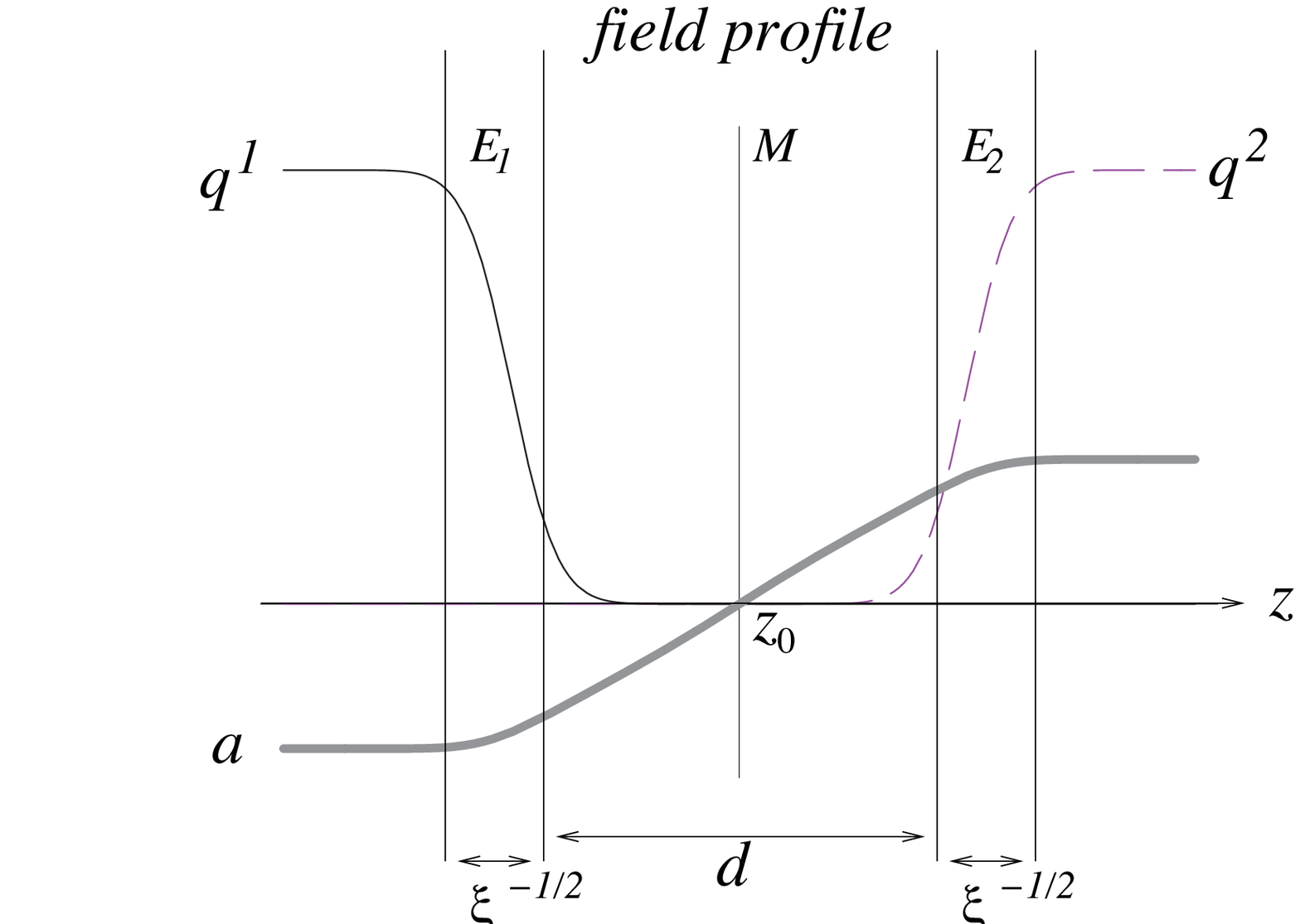}}
\caption{{\protect\small Internal structure of the domain wall: two edges
(domains $E_{1,2}$) of the width $\sim (g\protect\sqrt{\protect\xi})^{-1}$
are separated by a broad middle band (domain $M$) of the width $d$, see Eq.~(%
\protect\ref{d}).}}
\label{syfigthree}
\end{figure}

A crucial feature of the domain wall solution \cite{Shifman2002} is the
occurrence of a modulus --- a collective coordinate characterizing the wall
-- of an angular type. The origin of this collective coordinate $\alpha$ can
be explained as follows. The bulk theory at hand has in fact two U(1)
symmetries, corresponding to independent rotations of the first and second
flavors (see Table \ref{symmetries}). One of them is gauged, another global.
The global symmetry is spontaneously broken on the given wall solution.
That's why a massless phase field $\alpha$, a relative phase between two
flavors in the two respective vacua, is trapped on the wall. The $\alpha$
modulus supplements an obvious collective coordinate, the wall center $z_0$.


In what follows we will also consider the limit opposite to that in Eq.~(\ref%
{mggxi}), the so-called sigma-model limit \cite{GTT,Tw,GPTT},
\begin{equation}  \label{sigmamodelcondition}
\Delta m \ll g\sqrt{\xi} \,,
\end{equation}
In this limit the ``photonic" supermultiplet becomes heavier than that of
the second flavor field. Therefore, it can be integrated out. Then we are
left with the theory of (nearly) massless moduli $q^2,\,\tilde{q}_2$, which
interact through a nonlinear sigma model with the K\"ahler term
corresponding to the Eguchi--Hanson metric. The manifold parametrized by
these (nearly) massless fields is obviously four-dimensional. The vacua
discussed above lie at the base of this manifold. Therefore, in considering
the domain wall solutions in the sigma-model limit $\Delta m\to 0$ \cite%
{GTT,Tw,GPTT} one can limit oneself to the base manifold, which is, in fact,
a two-dimensional sphere. Classically, it is sufficient to consider the
domain wall in the CP(1) model deformed by a twisted mass term (related to a
nonvanishing $\Delta m$). This was first done in \cite{Tw}. A more general
analysis of the domain walls on the Eguchi--Hanson manifold can be found in
\cite{arai}. For our purposes it will be sufficient to limit ourselves to
the twisted-mass deformed CP(1) model.

The bosonic part of the Lagrangian of the sigma model is
\begin{equation}
\mathcal{L=}\frac{1}{2}\, \partial _{\mu }\vec{\varphi}\cdot \partial ^{\mu
} \vec{\varphi} -\frac{1}{2}\left(\Delta m\right)^{2}\left( \xi-
\varphi_{3}^{2}\right) ,\qquad \vec{\varphi}\cdot \vec{\varphi}=\xi\,.
\label{lagrangiansigmamodel}
\end{equation}
Using the stereographic projection (see e.g. the review paper \cite{nsvzr})
we can make the following change of coordinates:
\begin{equation}
u=\frac{\varphi _{1}+i\varphi _{2}}{\sqrt\xi -\varphi _{3}}\,.
\end{equation}
As a result, our sigma model takes the form
\begin{equation}
\mathcal{L}=\xi\left\{ \frac{\partial_\mu \bar{u}\,\, \partial_\mu u}{\left(
1+\bar{u}u\right) ^{2}} - (\Delta m)^{2}\frac{\bar{u}u}{\left( 1+ \bar{u}%
u\right) ^{2}}\right\}\,.  \label{lagrangiansigmamodelstereographic}
\end{equation}
Note that in the sigma-model limit only two fundamental parameters are left:
$\Delta m$ and $\xi$. Information about the coupling constant $g$ is lost.
In fact, this limit is sometimes called the ``strong coupling limit'' since
the condition (\ref{sigmamodelcondition}) can be reached by just sending the
coupling constant $g$ to infinity.

\begin{figure}[h!tb]
\epsfxsize=4.5cm \centerline{\epsfbox{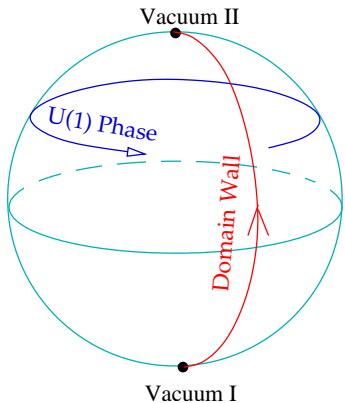}}
\caption{{\protect\small The sigma model target space, the domain wall and
its $U(1)$ modulus. The vacua $I$ and $I\!I$ correspond to $u=0$ and $%
u=\infty$, respectively. }}
\label{sigmamodel}
\end{figure}

We can readily obtain the BPS domain wall in the sigma-model approximation
too. In Fig.~\ref{sigmamodel} we display the sigma models target space, a
sphere $\mathbf{S}^2$. The two vacua --- the minima of the potential term
--- are the north and the south poles, corresponding to $u=\infty$ and $u=0$%
, respectively. The domain wall ``trajectory" is an arc interpolating
between the two poles. From Fig.~\ref{sigmamodel} the origin of the U(1)
modulus in this approximation is quite clear: The wall ``trajectory" can run
along any meridian. For a pedagogical review see \cite{shl}.

To get a quantitative description of the domain wall in the sigma-model
approximation we perform the Bogomol'nyi completion of the sigma-model
Lagrangian,
\begin{equation}
T=\int dz\, \xi\left\{ \left\vert \frac{\partial_{z}u\pm\Delta mu}{1+ \bar{u}%
u} \right\vert ^{2}\mp\Delta m \frac{\partial}{\partial z} \left(\frac{1 }{%
1+ \bar{u}u}\right) \right\}\,.  \label{bogcom}
\end{equation}
We observe that the BPS equation
\begin{equation}
\partial_z u =\mp u
\end{equation}
is holomorphic; therefore, the wall center modulus $z_0$ is complexified,
\begin{equation}
u = e^{\mp\Delta m(z-z_0) +i\alpha } \,,
\end{equation}
Any constant phase $\alpha$ is allowed. The full derivative term in Eq.~(\ref%
{bogcom}) implies that the wall tension
\begin{equation}
T_{\text{\textrm{w}}}=\xi| \Delta m |\,,
\end{equation}
in full accord with Eq.~(\ref{wten}).

The fact that each domain wall has two bosonic collective coordinates ---
its center and the phase --- in the sigma-model limit was noted in \cite{Tw}.


\begin{center}
\begin{table}[h!tb]
\begin{center}
\begin{tabular}{|c|c|c|c|c|}
\hline
\rule{0mm}{6mm}\!\! {\small Gauged U(1)} \!\! & \!$q^{1}\rightarrow
e^{i\alpha/2}q^{1}$ & \! $q^{2}\rightarrow e^{i\alpha/2}q^{2}$ & \! $\tilde
q_{1}\rightarrow e^{-i\alpha/2}\tilde q_{1}$ & \! $\tilde q_{2}\rightarrow
e^{-i\alpha/2}\tilde q_{2}$ \\ \hline
\rule{0mm}{6mm} \!\! {\small Global U(1)}\!\! & \! $q^{1}\rightarrow
e^{i\alpha/2}q^{1}$ & \!$q^{2}\rightarrow e^{- i\alpha/2}q^{2}$ & \! $\tilde
q_{1}\rightarrow e^{-i\alpha/2}\tilde q_{1}$ & \! $\tilde q_{2}\rightarrow
e^{ i\alpha/2}\tilde q_{2}$ \\ \hline
\end{tabular}%
\end{center}
\caption{{\protect\small Two U(1) symmetries of the bulk theory.}}
\label{symmetries}
\end{table}
\end{center}


Now let us describe the domain wall dynamics from another point of view:
that of the $2+1$ effective theory on the wall worldvolume. The wall
solution is characterized by two collective coordinates, the position of the
wall center $z_0$ and th phase $\alpha$. In the effective low-energy theory
on the wall they become scalar fields of the worldvolume(2+1)-dimensional
theory, $z_0 (t,x,y)$ and $\alpha (t,x,y)$, respectively. The target space
of the field $\alpha$ is $S_1$.

Let us make the wall collective coordinates $z_0$ and $\alpha $ (together
with their fermionic superpartners) slowly varying fields depending on $x_n$
($n=0,1,2$). Since $z_0 (x_n)$ and $\alpha (x_n)$ correspond to zero modes
of the wall, they have no potential terms in the worldsheet theory. It is
easy to derive the $z_0$ kinetic term,
\begin{equation}  \label{kinz0}
S_{2+1}^{z_0}= \frac{T_{\mathrm{w}}}{2} \, \int d^3 x \; (\partial_n z_0
)^2\, ,
\end{equation}
while the derivation of the normalization factor for the kinetic term for $%
\alpha$ is rather tedious, albeit straightforward \cite{Shifman2002},
\begin{equation}  \label{kinsig}
S_{2+1}^{\alpha}=\frac{\xi}{\Delta m}\, \int d^3 x \;\frac12 \, (\partial_n
\alpha)^2\, .
\end{equation}

As is well known from Polyakov's work \cite{P77}, the compact scalar field $%
\alpha (t,x,y)$ can be dualized into a (2+1)-dimensional Abelian gauge
field. Name\-ly,
\begin{equation}
F^{(2+1)}_{nm}=\frac{e^2}{2\pi} \, \varepsilon_{nmk}\, \partial^k \alpha\, ,
\label{21gaugenorm}
\end{equation}
where the (1+2)-dimensional coupling $e^2$ is
\begin{equation}
e^2=4\pi^2\, \frac{\xi}{\Delta m}\, .  \label{21coupling}
\end{equation}

Finally one gets the following effective low-energy theory of the moduli
fields on the wall:
\begin{equation}
S_{2+1}=\int d^3 x \, \left\{\frac{T_{\mathrm{w}}}{2}\,\, (\partial_n z_0
)^2+ \frac{1}{4\, e^2}\, (F_{nm}^{(2+1)})^2 +\mbox{fermion terms} \right\} .
\label{21theory}
\end{equation}

The emergence of the gauge field on the wall is easy to understand.
Let us start from a magnetic monopole in the bulk. Since in the bulk
the theory is fully Higgsed, the magnetic flux must be squeezed
inside an ANO flux tube \cite{ano} which starts on the monopole and
ends on the wall (it is oriented perpendicular to the wall). Inside
the wall the $q$ condensates vanish, the U(1) gauge group is
restored, and the magnetic flux spreads freely. In the Polyakov dual
language the magnetic flux lines spread inside the wall are the
electric flux lines on the wall. A vortex in $\alpha$ corresponds to
a charge source for the electric field on the wall.

The fermion content of the worldvolume theory is given by two
three-dimensional Majorana spinors, as is required by ${\mathcal{N}}=2$ in
three dimensions (four supercharges). The full worldvolume theory is a U(1)
gauge theory in $(2+1)$ dimensions, with four supercharges. The Lagrangian (%
\ref{21theory}) and the corresponding superalgebra can be obtained by
reducing four-dimensional ${\mathcal{N}}=1\,$ SQED (with no matter) to three
dimensions. The field $z_0$ in (\ref{21theory}) is the ${\mathcal{N}}=2\,$
superpartner of the gauge field $A_n$. To make it more transparent we make a
rescaling, introducing a new field
\begin{equation}
a_{2+1}=2\pi\xi\,z_0\,.  \label{az}
\end{equation}
In terms of $a_{2+1}$ the action (\ref{21theory}) takes the form
\begin{equation}
S_{2+1}=\int d^3 x \left\{ \frac{1}{2e^2} \left(\partial_n a_{2+1}\right)^2 +%
\frac{1}{4 e^2} \left( F_{mn}^{(2+1)} \right)^2 + \mbox{fermions}\right\}.
\label{pure}
\end{equation}
The gauge coupling constant $e^2$ has dimension of mass in three dimensions.

The lightest massive excitations of the wall have mass of order of the
inverse thickness of the wall $1/d$, see (\ref{d}). Thus the dimensionless
parameter that characterizes the coupling strength in the worldvolume theory
is $e^2 d$,
\begin{equation}
e^2 d=\frac{16\pi^2}{g^2}.  \label{duality}
\end{equation}
This can be interpreted as a feature of the bulk--wall duality: the weak
coupling regime in the bulk theory corresponds to strong coupling on the
wall and \emph{vice versa} \cite{Shifman2002}.


To conclude this section let us briefly summarize various limits considered
previously in this problem, and their range of validity.

\begin{description}
\item[Thin-edge approximation] :

In this approximation the mass of the scalar fields is much larger than the
photon mass, $|\Delta m| \gg g\sqrt{\xi}$. The domain wall has a three-layer
structure and its properties can be analyzed by virtue of very simple
equations.

\item[Sigma-model limit] :

In this approximation the mass of the scalar field is much smaller than the
photon mass, $|\Delta m| \ll g\sqrt{\xi}$. The heavy fields can be
integrated out, and what is left is a non-linear sigma model with the target
space $\mathbf{S}^2$ and a potential which leaves the north and the south
poles as the only vacua of the theory.

\item[Worldvolume effective action] :

Excitations of the wall moduli can be studied, in the low-energy limit, by
virtue of an effective Lagrangian on the domain wall worldvolume. Here by
low energies we mean energies much lower than $\Delta m$ and $g\sqrt{\xi}$.
\end{description}

We will deal with these limits wherever appropriate in the subsequent parts
of the paper.


\section{Fields Exited on the Wall} \label{fieldexited}

We now consider exited states of the domain wall. First switching on
a magnetic field and then an electric field.

\subsection{Constant magnetic flux inside the wall}

\label{cmfitw}

In Ref.~\cite{ds} it was noted that moduli fields in supersymmetric theories
have a new type of stable ``vacuum" solutions which preserve a part of
supersymmetry. In the case of the phase modulus this solution can be
presented as
\begin{equation}
\alpha = a \, x\,,  \label{lin}
\end{equation}
where $a$ is a constant of dimension of mass and $x$ is one of two
coordinates on the domain wall. It is clear that the solution (\ref{lin})
satisfies the equation
\begin{equation*}
\left( \frac{\partial^2}{\partial x^2} + \frac{\partial^2}{\partial y^2}%
\right) \alpha =0\,.
\end{equation*}

\begin{figure}[h!tb]
\epsfxsize=10cm \centerline{\epsfbox{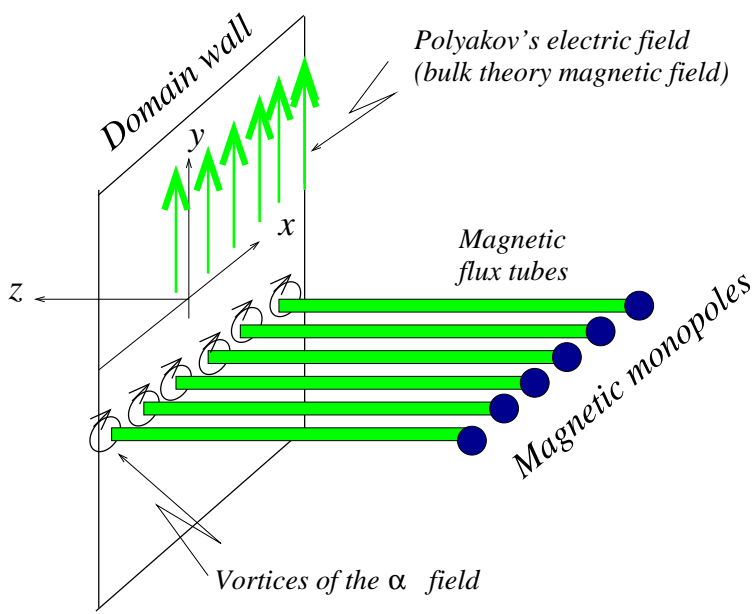}}
\caption{{\protect\small In the limit of vanishing distance between the
magnetic flux tubes (perpendicular to the wall) the field $\protect\alpha$
on the wall becomes linear in $x$. This is equivalent to a constant Polyakov
electric field on the wall surface aligned in the $y$ direction. From the
bulk point of view we have a constant magnetic field aligned in the $y$
direction propagating inside the wall parallel to its surface.}}
\label{mfie}
\end{figure}

\begin{figure}[h!tb]
\epsfxsize=4cm \centerline{\epsfbox{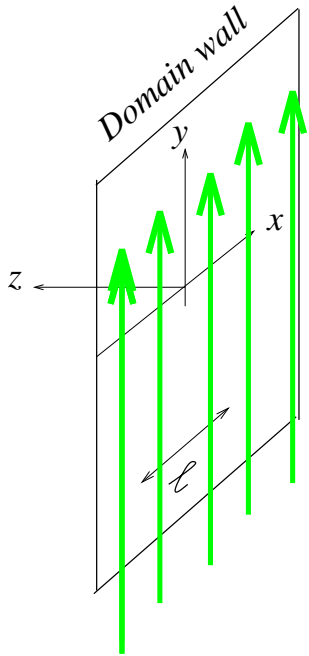}}
\caption{{\protect\small A domain wall with a constant magnetic flux
propagating inside the wall. The value of the magnetic flux per length $\ell$
in the $x$ direction is $2\, a\,\ell$. }}
\label{mfiep}
\end{figure}

In the Polyakov language this solution corresponds to a $F^{(2+1)}_{0y}$
component of the dual field strength tensor (\ref{21gaugenorm}). In the
bulk-theory language we describe in this way a constant magnetic field
inside the wall directed along the $y$ axis. Figure~\ref{mfie} illustrates
the validity of this interpretation. The bulk part is depicted only with the
purpose of making clear the physical basis of the construction. After this
is done, one can eliminate it and consider an infinite wall stretched in the
$xy$ plane with a constant magnetic flux propagating inside the wall (Fig~%
\ref{mfiep}). The value of the flux per length $\ell$ in the $x$ direction
is $2\, a\,\ell$. The factor 2 here is in one-to-one correspondence with way
we define the electric charge in Eq.~(\ref{nab}), i.e. the coefficient $%
\frac{1}{2}$ in front of $A_\mu$. In this definition the magnetic flux of
the magnetic monopole is $4\pi$.

The absolute value of the magnetic field inside the wall is
\begin{equation}
|\vec B| = 2 d^{-1}\, a\,,
\end{equation}
where for definiteness $a$ is assumed to be positive.

The topological defect depicted in Fig.~\ref{mfiep} is stable --- the vacua
on the right- and left-hand sides of the wall are distinct. The tension of
this defect is
\begin{equation}
T_{\mathrm{w\, B}}=\xi \, \Delta m\left[ 1+\frac{a^2}{2(\Delta m)^2} \right]%
\,,  \label{wtenp}
\end{equation}
where the subscript B corresponds to the presence of the magnetic field $B$
inside the wall. We have to assume that $a/\Delta m \ll 1$. We will discuss
later how this limitation can be lifted.

In the case of the infinite wall, the magnetic flux is not quantized.
However, the flux quantization does take place if we compactify the
horizontal direction, making a cylinder from the wall at hand. That's what
we will do below.


\subsection{Time dependence of $\protect\alpha$: what does it
mean?}

\label{mean}

From the point of view of the classical equation of motion and stability a
linear function in $t$ is as good a solution as (\ref{lin}). Assume that
\begin{equation}
\alpha = \omega \, t\,,  \label{lint}
\end{equation}
where $\omega$ is a constant of dimension of mass. In the Polyakov language
this means that an $xy$ component of the dual $(2+1)$-dimensional field is
generated,
\begin{equation}
F^{(2+1)}_{12}=2\pi\, \frac{\xi\, \omega}{\Delta m} \,.  \label{pmag}
\end{equation}

\begin{figure}[h!tb]
\epsfxsize=8cm \centerline{\epsfbox{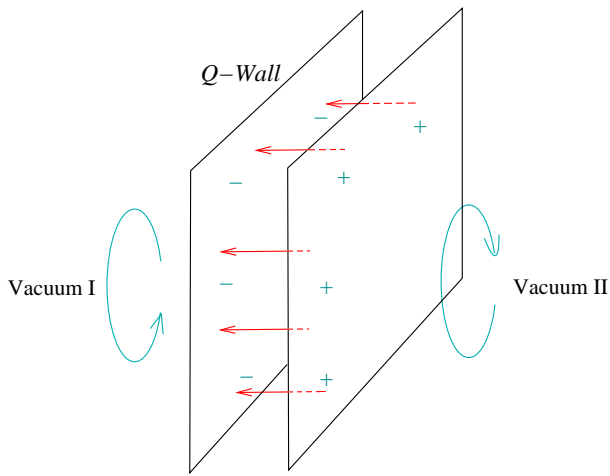}}
\caption{{\protect\small Q charged domain wall. The electric field inside
the wall (it is perpendicular to the wall surface) is denoted by arrows. }}
\label{mfiepp}
\end{figure}

The $xy$ component is the Polyakov dual magnetic field (which is a scalar
with respect to rotations in the $xy$ plane). In the bulk theory this
corresponds to an electric field perpendicular to the wall surface, see Fig.~%
\ref{mfiepp}. If $\omega\neq 0$, then the tension of the wall with the
electric and magnetic fields inside is
\begin{equation}
T_{\mathrm{w\, B\,E}}=\xi \, \Delta m\left[ 1+\frac{a^2}{2(\Delta m)^2} +
\frac{\omega^2}{2(\Delta m)^2} \right]\,,  \label{wtenpp}
\end{equation}
where the subscript E reminds us of the electric field $E$ inside the wall.
Equation (\ref{wtenpp}) implies that the absolute value of the electric
field inside the wall is
\begin{equation}
|\vec E| = 2 d^{-1}\, \omega\,.
\end{equation}

So far we focused on the worldvolume theory on the wall and interpreted the
``vacua" with $\alpha$ linearly dependent on $t,x$ in terms of $\vec B$ and $%
\vec E$ inside the wall. Let us ask ourselves about the picture of this
phenomenon viewed from outside, from the bulk.

The bulk theory (\ref{n2sqed}) has two U(1) symmetries: one global, one
local. The local U(1) is realized in the Higgs mode, i.e. there is a charge
condensate in the bulk and no massless gauge bosons. This is the reason why
the electric field inside the wall must be perpendicular to its surface. The
charge $Q$ corresponding to the conserved global U(1) which emerges after
the theory is Higgsed in the bulk can be chosen on the wall surface at will
as long as it does not change with time. If $\omega \neq 0$ the domain wall
carries a nonvanishing density of $Q$,
\begin{equation}
Q/\mathrm{Area} \, =\, \frac{\xi\, \omega}{2\,\Delta m}\,,
\end{equation}
where $\omega $ is assumed to be positive. $Q$-charged solitons were
invented by Coleman, who considered a particular type \cite{Coleman:1985ki}
currently known as $Q$ balls. Generalizations were discussed in \cite%
{Leese,Abraham:1991ki}.

Elementary excitations of the vacuum carry a nonvanishing $Q$ charge too.
For instance, to the right of the wall, far away from it (Fig.~\ref%
{syfigthree}) one can always require $q^2$ to be real. The quantum of $q^1$
carries a phase, however, and, as a result, will have $Q =\pm 1$. The
question is what prevents $Q$ from leaking from the wall into the bulk
through particle emission.

The answer to this question seems to be rather straightforward. A unit
charge inside the wall edge is lighter than a particle with the same charge
propagating in the vacuum. In other words, the work function is high.
Indeed, the area of the segment of the wall with the unit U(1) charge is
\begin{equation}
A\sim \frac{\Delta m}{\xi\, \omega}\,.  \label{fbe}
\end{equation}
The excess of energy corresponding to this segment is
\begin{equation}
\Delta E = \Delta T\, A \sim \omega\,,  \label{fbee}
\end{equation}
where we used Eqs. (\ref{wtenpp}) and (\ref{fbe}). This is to be compared
with the mass of the $Q=1$ particle, $\Delta m$ (Sect.~\ref{bulkth}).
Stability is guaranteed provided
\begin{equation}
\omega \ll \Delta m\,.  \label{fbeee}
\end{equation}

Stability of solitons of this type in theories with a conserved U(1) charge
is well studied in supersymmetric models. For a recent analysis of
two-dimensional ${\mathcal{N}}=2$ sigma model with twisted mass see~\cite%
{SVZ}. In this model there is a domain in the parameter space in which $Q$%
~kinks cannot emit elementary $Q$ charged excitations because such emission
is energetically forbidden. A curve of marginal stability separates this
domain from that where the tower of $Q$ charged kinks decays.



A remark is in order here regarding ``$Q$ charging" from the standpoint of
the bulk theory. Examining the Lagrangian (\ref{n2sqed}) we see that the $Q $%
~kink is a soliton with \emph{four} excited fields: three scalar fields $q_1$%
, $q_2$, $a$, and the time component of the gauge potential $A_t$. Table~\ref%
{profilesQkink} gives a summary of the corresponding profile functions, with
the appropriate boundary conditions.

\begin{center}
\begin{table}[h!tb]
\begin{center}
\begin{tabular}{|c|c|c|}
\hline
\rule{0mm}{6mm} & $x_{3}\rightarrow -\infty $ & $x_{3}\rightarrow +\infty $
\\ \hline
\rule{0mm}{6mm} $q_{1}\left( x_{3};t\right) $ & $\sqrt{\xi} e^{i \frac{\omega%
}{2} t}$ & $0$ \\ \hline
\rule{0mm}{6mm} $q_{2}\left( x_{3};t\right) $ & $0$ & $\sqrt{\xi} e^{-i
\frac{\omega}{2} t}$ \\ \hline
\rule{0mm}{6mm} $a \left( x_{3}\right) $ & $-\sqrt{2} m_{1}$ & $-\sqrt{2}
m_{2}$ \\ \hline
\rule{0mm}{6mm} $A_{t}\left( x_{3}\right) $ & $\omega $ & $-\omega $ \\
\hline
\end{tabular}%
\end{center}
\caption{{\protect\small Profile functions relevant to the $Q$-kink soliton
in the bulk theory. Note that $x_3\equiv z$. }}
\label{profilesQkink}
\end{table}
\end{center}

$Q$ charging is attained in two steps. First we take the domain wall and
give it a global U(1) phase rotation $e^{i\omega t}$. However, this step
alone would result in a divergence in the wall tension due to the time
derivatives of the type $\partial _{t}\,q$. The second step is to compensate
the time dependence of $q$'s far away from the wall by introducing a gauge
potential $A_t$.

A subtle point here is that the U(1) gauge transformation is different from
the global U(1). They can be identified, however (in two distinct ways
depending on whether we are to the left or to the right from the wall).
Namely, at $x_3 \to -\infty$ we must have $A_t\to \omega$ while at $x_3 \to
+\infty$ we must have $A_t\to -\omega$. The sign of the gauge potential is
different at two extremities. Step two thus cures the divergence of the wall
tension but it leaves a trace. $A_t$ is in fact pure gauge only far away
from the wall. In wall's interior an electric field is generated.

The $Q$ wall in the full bulk theory can be analyzed with more precision. We
refer the reader e.g. to Ref.~\cite{tongreview} for the Bogomol'nyi
completion and derivation of the first-order differential equations.

Below we use the bulk theory picture in order to generalize Eq.~(\ref{wtenpp}%
) and other similar expressions to larger values of the $Q$ charge.


\subsection{$Q$ wall in the thin-edge approximation}

\label{qwtea}

In the thin-edge approximation the $Q$ wall can be viewed as an
electrostatic capacitor. Let us assume there is a positive charge density $%
+\sigma$ on one edge (say, on the left) and a negative charge density $%
-\sigma$ on the other edge.\,\footnote{%
In our conventions the electromagnetic Hamiltonian ${\mathcal{H}}=
(2g^2)^{-1}\left(\vec E^2 +\vec B^2 \right)$. Then the Gauss' theorem
implies that the charge $= g^{-1}\int_{{\mathcal{S}}_R}d^2 S_i {E}_i$. The
electric field $E$ in the capacitor is $|E| = \sigma$. The minimal charge of
the quantum of the $q$ field is 1/2, see Eq.~(\ref{nab}). Thus, in our
normalization the unit of the U(1) charge is in fact $g^2/2$.}

Inside the capacitor, between the two edges, we have an electric field $%
E_{z}=\sigma$. The wall tension is now given by a sum of three terms,
\begin{equation}
T_{Q-\mathrm{wall}}\left( d \right) =\frac{2 (\Delta m)^{2} }{g^{2} d}+
\frac{g^{2}\xi^{2}}{8} d+\frac{\sigma^{2}}{2g^{2}}d\,.  \label{3terms}
\end{equation}
Equation (\ref{3terms}) replaces Eq.~(\ref{2terms}) where the electric field
inside the wall is not included. Inclusion of the electric field is
equivalent to the substitution
\begin{equation}
\xi^2 \to \xi^2 +\frac{4\sigma^2}{g^4}\,.
\end{equation}
Minimizing with respect to $d$ in the same way as above we obtain
\begin{equation}
d_{Q-\mathrm{wall}} =\frac{4 \Delta m}{g^2\,\xi } \left(1+ \frac{4\sigma^2}{%
g^4\,\xi^2} \right)^{-1/2}\,,  \label{segt}
\end{equation}
and
\begin{equation}
T_{Q-\mathrm{wall}}= \Delta m\,\xi \left(1+ \frac{4\sigma^2}{g^4\,\xi^2}
\right)^{1/2}\,.  \label{segtp}
\end{equation}
One can see that Eq.~(\ref{2terms}) presents a small-$\sigma$ expansion of
Eq.~(\ref{segtp}) provided one identifies
\begin{equation}
\omega = 2\,\sigma\,\frac{ \Delta m}{g^2\,\xi}\,.  \label{segtpp}
\end{equation}

We already know that at small $\omega$ the $Q$ charged wall is stable with
respect to ``leakage" of the U(1) charge outside in the form of emission of
quanta of the $q$ fields. Now we can check that this remains valid for large
$Q$ charges as well. In fact, in the limit $\sigma\to\infty$ the $Q$ wall
becomes marginally stable (Fig.~\ref{qkinktension}).

Indeed, at large $\sigma$ the wall tension becomes linear in $\sigma$,
\begin{equation}
T_{Q-\mathrm{wall}} = \frac{2}{g^2}\, \Delta m \,\sigma \,.
\end{equation}
Let us pick up an area $A$ such that a change $\delta \sigma$ on this area
could produce a quantum of the unit charge. Given our normalization $%
A\,\delta\sigma =g^2/2$. The corresponding variation of the energy of the
wall is
\begin{equation}
\delta E = A\, \delta T_{Q-\mathrm{wall}} = A\, \frac{2}{g^2}\, \Delta m
\,\delta\sigma =\Delta m\,.
\end{equation}
This is precisely the mass of a single $q$ quantum.


\subsection{$Q$ wall in the sigma-model limit}

\label{qwsml}


Alternatively we can consider the $Q$ wall in the sigma model approximation.
To obtain the wall solution in the sigma model we perform the following
completion procedure:
\begin{eqnarray}
T & =& \int dz\, \xi\left\{ \frac{\partial_{t}\bar{u} \, \partial_{t}u}{%
\left( 1+ \bar{u}u\right) ^{2}}+ \frac{\partial_{z}\bar{u} \, \partial_{z}u}{%
\left( 1+\bar{u}u\right) ^{2}}+ (\Delta m)^{2}\frac{\bar{u}u}{\left( 1+\bar {
u}u\right) ^{2}} \right\}  \notag \\[3mm]
& = & \int dz\, \xi\left\{ \frac{\left[ \partial_{t}u\pm i ( \Delta m)
\left( \sinh b\right) u\right] \left[ \partial_{t}\bar{u}\pm i(\Delta m)
\left( \sinh b\right) \bar{u}\right] }{\left( 1+\bar{u}u\right) ^{2}}\right.
\notag \\[3mm]
&+& \left\vert \frac{\partial_{z}u\pm (\Delta m) \left( \cosh b\right) u}{1+%
\bar{u}u} \right\vert ^{2} \pm i (\Delta m) \left( \sinh b\right)
\partial_{t} \left[\frac {1}{ 1+\bar{u}u }\right]  \notag \\[3mm]
&\pm & \left.(\Delta m) \left(\cosh b\right) \partial_{z} \left[\frac {1}{ 1+%
\bar{u}u }\right]\right\} \,,
\end{eqnarray}
where $b$ is a constant. The solution takes the form (for the lower choice
of the sign)
\begin{equation}
u= e^{i(\Delta m) \left( \sinh b\right) t}\, e^{(\Delta m) \left(\cosh
b\right) (z-z_0)}\,,
\end{equation}
while the tension
\begin{equation}
T_{\sigma-\mathrm{model}}=\xi (\Delta m) \left(\cosh b\right)\,.
\end{equation}
To compare with our previous results it is convenient to normalize $b$ as
\begin{equation}
\omega= (\Delta m )\left(\sinh b\right)
\end{equation}
where the frequency $\omega$ was introduced earlier. Then
\begin{equation}
T_{\sigma-\mathrm{model}}=\xi\, \sqrt{(\Delta m)^{2}+\omega^{2}}\,,
\end{equation}
which at small $\omega$ reduces to the $\omega$ dependent part of Eq.~(\ref%
{wtenpp}). Moreover, the wall solution takes the form
\begin{equation}
u= e^{i\omega t}\, e^{\sqrt{(\Delta m)^{2}+\omega^{2}}\, (z-z_0)}\,.
\end{equation}

\begin{figure}[h!tb]
\epsfxsize=12cm \centerline{\epsfbox{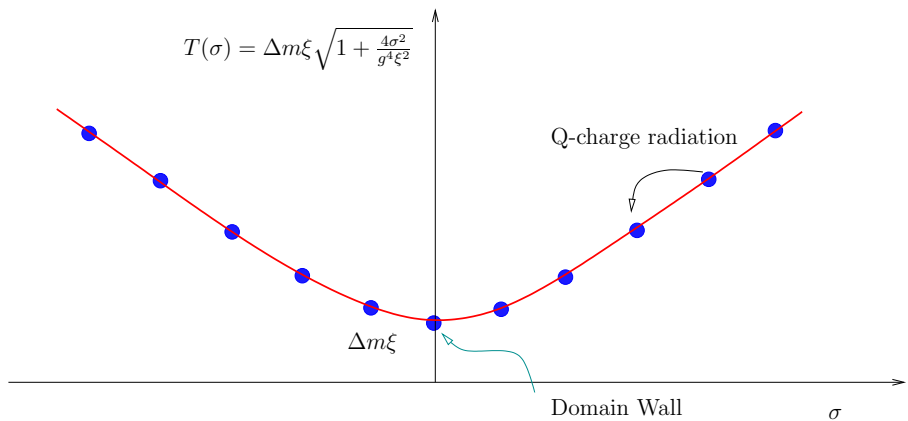}}
\caption{{\protect\small The Q wall tension vs. $\protect\sigma$. In our
normalization the elementary unit of the U(1) charge is $\frac{g^2}{2}$. The
Q wall is stable with respect to radiation of $q$ quanta. It becomes
asymptotically marginally stable at $\protect\sigma \to \infty$. }}
\label{qkinktension}
\end{figure}

One can readily see that we get the same expression for the $Q$ wall tension
as in the thin-edge approximation (cf. Eq.~(\ref{segtpp})). Therefore, the
same stability argument applies. The marginal stability at large $Q$ is
clearly seen from Fig.~\ref{qkinktension}.

\section{$Q$ cylinder}

\label{cylinder}

As was mentioned, the basic wall depicted in Fig.~\ref{syfigthree} is
absolutely stable as long as it is flat and its area is infinite. This
stability does not require electric or magnetic fluxes which may or may not
be introduced with no impact on stability.

Our strategy is to start from this flat wall, and bend/fold it in an
appropriate way to get solitons with a more complicated geometry without
loosing stability. A natural starting point is cylindrical geometry.

\begin{figure}[h!tb]
\epsfxsize=12cm \centerline{\epsfbox{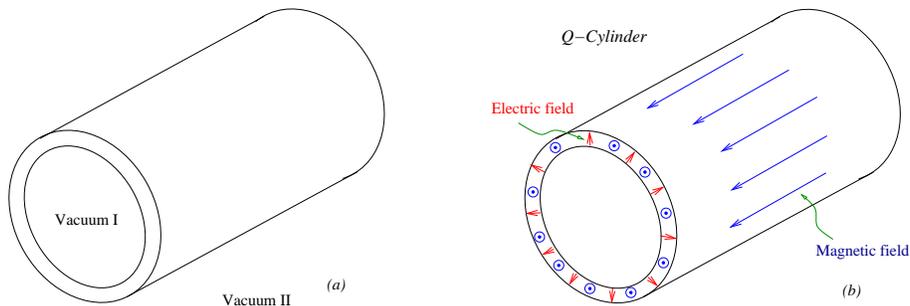}}
\caption{{\protect\small Folding the domain wall. (a) Making a cylinder from
the basic (``empty") wall; (b) Magnetic and electric fields inside the wall.}
}
\label{qcy}
\end{figure}

If we try to fold the ``empty" wall (no fluxes) in the form of a cylinder
(Fig.~\ref{qcy}\,a) it becomes unstable. Nothing prevents it from shrinking
in the radial direction until the vacuum $I$ is completely ``squeezed out"
so that the wall-cylinder decays into a large number of elementary
excitations in the vacuum $I\!{I}$. Of course, if the radius of the cylinder
is much larger than the thickness of the wall --- and we will assume this to
be the case --- the process of shrinkage will be slow until the field
configuration at hand becomes a ``thin cylinder,'' i.e. the one whose radius
is only slightly larger than the wall thickness.

Can one stabilize the cylinder? The first thought that comes to one's mind
is to let the magnetic flux propagate inside the cylinder in the direction
of its axis. Then disintegration of the cylinder via decays into elementary
excitations in the vacuum $I\!{I}$ is impossible since the magnetic flux is
conserved. However, the magnetic flux conservation does not prevent the
cylinder from passing into the ANO strings~\cite%
{ano} with the same value of the total magnetic flux, plus elementary
excitations. Whether or not the cylinder with the magnetic field trapped
inside the wall is stable with respect to such decays is a dynamical
question. Note that in the cylindrical geometry the magnetic flux gets
quantized.

Assume that the radius of the cylinder is very large so that locally a flat
wall approximation is applicable. Then the phase field $\alpha$ will
continue to provide an adequate description of the worldsheet dynamics.
However, globally the change of $\alpha$ in the compactified direction must
be of the form $2\pi\, n$ with an integer $n$. If the circumference of the
cylinder is $L$, this implies
\begin{equation}
\alpha = 2\, \pi \, n \, \frac{x}{L}\,.  \label{akx}
\end{equation}
Correspondingly, the magnetic flux trapped inside the wall is $4\,\pi\, n$,
i.e. that of $n$ magnetic monopoles. At $n=1$ the flux is the same as the
ANO string. The cylindrical wall will be stable against decays into the ANO
string(s) plus elementary excitations provided the $n=1$ cylinder tension is
less than that of the ANO string. Unfortunately, this is not the case. For
instance, in the sigma-model approximation one can readily get that the
ratio of the above tensions is $\sqrt 2$. The cylindrical wall with the
magnetic flux is heavier than the ANO strings with the same flux.
Intuitively this seems natural since the internal structure of the
cylindrical wall is more complicated than that of the ANO string.

Thus, the simplest attempt to stabilize the wall-cylinder fails. This is not
the end of the story, however. In addition to the magnetic field inside the
wall, one can $Q$-charge it, i.e. introduce an electric field in the radial
direction, as in Fig.~\ref{qcy}\,b. For sufficiently large radius of the
cylinder we still can write
\begin{equation}
\alpha = 2\pi \,n\, \frac{x}{L} + \omega t\,,
\end{equation}
where $x$ is the coordinate perpendicular to the cylinder axis (see Fig.~\ref%
{coor}).

\begin{figure}[h!tb]
\epsfxsize=5cm \centerline{\epsfbox{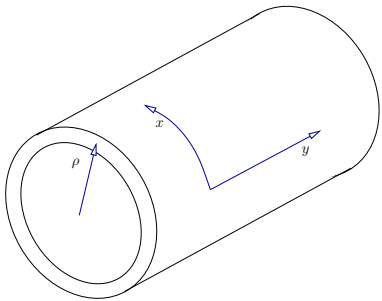}}
\caption{{\protect\small Geometry of the cylindrically folded wall. }}
\label{coor}
\end{figure}

The non-vanishing value of the $Q$ charge does make the wall-cylinder stable
since the ANO strings are $Q$ neutral, which precludes the decay of the $Q$
charged wall-cylinder into the ANO strings plus elementary excitations.
Elementary excitations are $Q$ charged, generally speaking, but as was
discussed in Sect.~\ref{mean}, the leakage of the wall $Q$ charge into
elementary excitations is energetically forbidden, at least as long as the
wall is nearly flat.

The shrinkage of the cylinder in the radial direction due to the wall
tension is stabilized by the magnetic field inside the wall. Indeed, let $%
\rho$ denote the cylinder radius (Fig.~\ref{coor}) and $4\pi \, n$ the
magnetic flux. Then the magnetic field
\begin{equation}
|\vec B | \sim \frac{n}{\rho\, d}\,,
\end{equation}
where $d$ is given in Eq.~(\ref{d}). Thus, the energy stored in the magnetic
field per unit length of the cylinder is
\begin{equation}
{\mathcal{E}}_B \sim \frac{n^2}{g^2\, d\,\rho}\,.  \label{ebcyl}
\end{equation}

A more detailed consideration of the $Q$ cylinder from the bulk-theory
standpoint is given in Sect.~\ref{qcbstp} where we determine the critical
values of $\rho$ and $d$ in terms of $Q$ and $n$ and formulate the condition
of ``thin cylinder walls" implying $\rho\gg d$.

\subsection{$Q$ cylinder from the bulk-theory standpoint}

\label{qcbstp}

Here we will parallel our discussion of the $Q$ wall in the thin-edge
approximation carried out in Sect.~\ref{qwtea} extending it to the $Q$
cylinder. The method we use is very similar to the one used in \cite%
{Bolognesi:2005,Bolognesi:2005rj} to study the large-$n$ limit of the
Abrikosov--Nielsen--Olesen vortex. As in Sect.~\ref{cylinder} we will assume
the radius of the cylinder $\rho$ to much larger than the wall thickness $d$
(which, in turn, is much larger than the thickness of the edge layers, to be
set to zero in our approximation). As in the $Q$ wall case, there are four
distinct contributions to the $Q$ cylinder tension: (i) the loss of energy
due to the Coulomb phase inside the cylinder wall; (ii) kinetic energy of
the $a$ field, (iii) the energy of the magnetic field inside the cylinder
wall; and (iv) the energy of the electric field. The corresponding
expressions have a somewhat different form. Let us discuss them one by one.
Our starting point will be Eq.~(\ref{3terms}).

(i) The loss of energy due to the Coulomb phase inside the cylinder wall per
unit length of the cylinder can be obtained from the second term in Eq.~(\ref%
{3terms}),
\begin{equation*}
\frac{\pi}{4}\, g^2 \xi\, \rho\, d\,.
\end{equation*}

\noindent (ii) The kinetic energy of the $a$ field per unit length of the
cylinder can be obtained from the first term in Eq.~(\ref{3terms}),
\begin{equation*}
\frac{4\pi}{g^2}\,(\Delta m)^2\, \frac{\rho}{d}\,.
\end{equation*}

\noindent (iii) The magnetic field energy per unit length is (see Eq.~(\ref%
{ebcyl}))
\begin{equation*}
\frac{4\pi n^2}{g^2}\, \frac{1}{\rho\, d}
\end{equation*}

\noindent (iv) Finally, to obtain the electric field energy per unit length
we must define $Q$ per unit length of the cylinder. In Eq.~(\ref{3terms}) we
introduced $\sigma$, the density of the U(1) charge per unit area (in the
units of $g^2/2$). It is convenient then to introduce $Q$ as
\begin{equation}
\sigma\, (2\pi\rho )= Q\,\frac{g^2}{2}\,.
\end{equation}
The electric field energy per unit length takes the form
\begin{equation*}
\frac{1}{16\pi}\,g^2\,Q^2\frac{d}{\rho}\,.
\end{equation*}
As a result, the total tension of the $Q$ cylinder (per its unit length) is
the following function of $\rho$ and $d$:
\begin{equation}
T_{Q-\mathrm{cylinder}} (\rho,d)= \underset{\mathrm{Coulomb\ phase}}{
\underset{}{\frac{g^{2}\xi^{2}\pi}{4} \rho d}}+ \underset{\mathrm{Kinetic}
\text{\textrm{\ }}\ a }{\underset{}{\frac{4\pi (\Delta m)^{2}}{g^{2}} \frac{%
\rho}{d }}} + \underset{\mathrm{Magnetic\ field}}{\underset{}{\frac{4\pi
n^{2} }{g^{2}}\frac{1}{\rho d}}}+ \underset{\mathrm{\ Electric\ field}}{%
\underset{}{\frac{Q^{2}g^2}{16\pi } \frac{d }{\rho }}} \,.
\label{tension Qlump formula}
\end{equation}

\vspace{2mm}

We do minimization in two steps. First we minimize the ``Coulomb" and the
magnetic field terms to obtain the product $\rho d$ at the minimum. Then we
minimize the kinetic-$a$ and the electric field terms to obtain the ratio $%
\rho / d$ at the minimum. In this way we arrive at (the asterisk denotes the
values at the minimum)
\begin{equation}
\left(\rho\,d \right)_* =\frac{4n}{g^2\,\xi}\,, \qquad \left(\frac{\rho}{d }%
\right)_* =\frac{Q\,g^2 }{8\pi (\Delta m) }\,.  \label{randd}
\end{equation}
Inserting these critical values in Eq.~(\ref{tension Qlump formula}) we get
the tension of the $Q$ cylinder,
\begin{equation}
T_{Q-\mathrm{cylinder}} =2\pi \xi n+\ Q\,(\Delta m)\,.  \label{qcylbt}
\end{equation}
For consistency we must require $\left({\rho}/{d }\right)_*\gg 1$.

\subsection{Implementation of the $Q$ cylinder in the sigma-model
limit}

\label{implqc}

It turns out that in the sigma-model limit an analog of the $Q$ cylinder was
discovered and studied long ago, see Ref.~\cite{Leese} where it goes under
the name of a $Q$-lump. We will reserve the name $Q$ lump for
(2+1)-dimensional theory in which this soliton must be considered as a
particle. In (3+1)-dimensional theory we are interested in Leese's soliton
can be called $Q$ string. The $Q$ string is characterized by a topological
(instanton) number $n$ and the global U(1) charge per unit length $Q$. In
essence, it is a combination of the Belavin--Polyakov instanton \cite{BP} in
the two-dimensional CP(1) model with $Q$ charging. What remains to be done
is to identify the instanton quantum number, the topological charge $n$,
with the magnetic flux in the full theory (in our normalization the magnetic
flux is $4\pi n$). We will see shortly that this identification is correct.
Thus, in the sigma-model limit the magnetic flux of the $Q$ cylinder becomes
related to the homotopy group $\pi_2(S^2)$. In other words, it is related to
the number of times the plane is mapped onto the target-space sphere. A
special feature of the Leese-type $Q$ charged soliton is that the soliton
solution approaches the vacuum asymptotics in a power-like rather than
exponential manner.

The conventional Belavin--Polyakov instanton \cite{BP} (for a review see
\cite{nsvzr}) has two moduli: the size and the U(1) phase. The size is a
modulus since the CP(1) model considered by Belavin and Polyakov is
classically scale invariant.

Addition of the twisted mass in the sigma model and associated with it
potential brakes the scale invariance. The instanton thus tends to shrink to
zero size and become singular. A way to prevent this collapse is to $Q$
charge the soliton, i.e. to give a time-dependent rotation to the U(1)
phase. Then the size modulus gets fixed in terms of $Q$.

To obtain the exact solution for the $Q$ string in the sigma model we
perform the following Bogomol'nyi completion:
\begin{eqnarray}
T_{Q-\mathrm{string}} & =& \xi\, \int d^{2}x\, \left[ \left\vert \frac{%
\partial_{t}u\pm i (\Delta m) u}{1+ \bar{u}u}\right\vert ^{2}+ \frac{1}{2}
\left\vert \frac{\partial_{i}u\pm i\epsilon_{ij}\partial_{j}u}{1+ \bar{u}u}
\right\vert ^{2}\right.  \notag \\[4mm]
& \pm& \left. i\, (\Delta m )\, \frac{\bar{u}\partial_{t}u-u\partial_{t}\bar{%
u}}{\left( 1+\bar{u}u\right)^{2}} \mp \frac{ i}{2}\, \epsilon_{ij}\frac{%
\partial_{i}\bar {u} \partial_{j}u-\partial_{j}\bar{u}\partial_{i}u}{\left(1+%
\bar{u}u\right)^{2} } \right]\,.  \label{qcylb}
\end{eqnarray}
The U(1) charge per unit length is
\begin{equation}
Q=\xi\, \int d^{2}x\, i\, \frac{\bar u\,\dot{u} - u\dot{\bar u}}{\left( 1+%
\bar{u}u\right) ^{2}}\,,  \label{qchd}
\end{equation}
while the last term is proportional to the instanton invariant
\begin{equation}
n = -\frac{i}{4\pi}\, \epsilon_{ij} \frac{\partial_{i}\bar {u} \partial_{j}u
-\partial_{j}\bar{u}\partial_{i}u}{\left( 1+\bar{u}u\right) ^{2}}\,.
\label{inn}
\end{equation}
For 1/2 BPS-saturated $Q$ string the first line in Eq.~(\ref{qcylb}) must
vanish. Choosing the lower signs the saturating solution can be written as
\begin{equation}
u(z,t)=u_0(z)\, e^{i(\Delta m) t}\,,  \label{bpip}
\end{equation}
where we introduce the complex variable $z= x_1 - i x_2$. The equation for $%
u_0(z)$ is then the Belavin--Polyakov instanton equation (self-duality
equation) which has a solution
\begin{equation}
u_0(z)=\left( \frac{c}{z}\right)^n \, ,  \label{bpi}
\end{equation}
where $c$ is a constant, while the power $n$ coincides with the instanton
number in Eq.~(\ref{inn}). Calculating $Q$ for the solution (\ref{bpip}), (%
\ref{bpi}) we find
\begin{equation}
Q=\xi \, (\Delta m)\, |c|^2 \, \frac{2\pi^2 }{n^2} \,\left(\sin\frac{\pi}{n}
\right)^{-1}\,.  \label{bpiq}
\end{equation}
For the minimal winding, $n=1$, the charge density $Q$ is infinite: in this
case Eq.~(\ref{qchd}) is logarithmically divergent. This is due to the fact
that the Leese soliton approaches the vacuum field in a power-like manner.
Only at large $n$ it becomes reminiscent of a thin-wall cylinder. And even
at large $n$ the relation with the worldvolume description of Sect.~\ref%
{cylinder} is not straightforward. Indeed, in the worldvolume description
the phase $\alpha$ could be any linear function of $t$, while for the $Q$
string at hand the phase is $(\Delta m) t$, with the unambiguously fixed
coefficient. The charge density $Q$ can still be arbitrary since it
quadratically depends on the instanton size $|c|$, see Eq.~(\ref{bpiq}). If $%
Q$ is fixed, so is the $Q$ string transverse size. This is not a feature of
the worldvolume description.

After these remarks we can finally give the expression for the 1/2
BPS-saturated $Q$~string tension,
\begin{equation}
T_{Q-\mathrm{string}} = 2\pi \xi n + \left(\Delta m\right) Q\,.  \label{qstt}
\end{equation}
This formula confirms our interpretation of the instanton number $n$ as the
magnetic flux in the microscopic theory.

In this model, thanks to the underlying supersymmetry, there is a
considerable amount of exact analytic information for studying the important
issue of $Q$ string stability. They are stable against decay into elementary
quanta of the $u$ field perturbed around the vacuum.

The solution (\ref{bpi}) is a limiting case of the $n$-instanton solution,
which is known in the general case. A generic solution with the arbitrary
instanton number $n$ can be written as
\begin{equation}
u(z,t) =\left( \sum_{i=1}^{n} \frac{c_{i}}{z-z_i}\right) \, e^{i\Delta m
t}\,,  \label{genn}
\end{equation}
where the moduli $c_i$ are complex numbers subject to the constraint
\begin{equation}
\sum_i^n c_i =0 \,.  \label{cmodc}
\end{equation}
This polygonal relation ensures finiteness of the $Q$ charge (per unit
length) and $T_{Q-\mathrm{string}}$. Thus, for any given finite value of the
$Q$ charge the dimension of the moduli space of the $Q$ string is $\nu =4n-2$%
.


\subsection{Comparing the bulk-theory and sigma model approximations}

\label{cbtsma}

It is instructive to compare results for the $Q$ cylinder obtained in
various approximations. First, we observe with satisfaction that the
expressions for the tension, Eqs.~(\ref{qcylbt}) and Eq.~(\ref{qstt}),
coincide. It is instructive to make one step further. Let us discuss the
moduli space of the $Q$ cylinder. The sigma model analysis demonstrates that
the $Q$ string in the topological sector $n$ has a moduli space of solutions
of dimension $4n-2$. For our consideration to be valid we must assume $n\gg
1 $. In this limit we deal with an infinite-dimensional moduli space in the
sigma-model description. Is there a trace of this phenomenon in the
bulk-theory description?

\begin{figure}[h!tb]
\epsfxsize=12cm \centerline{\epsfbox{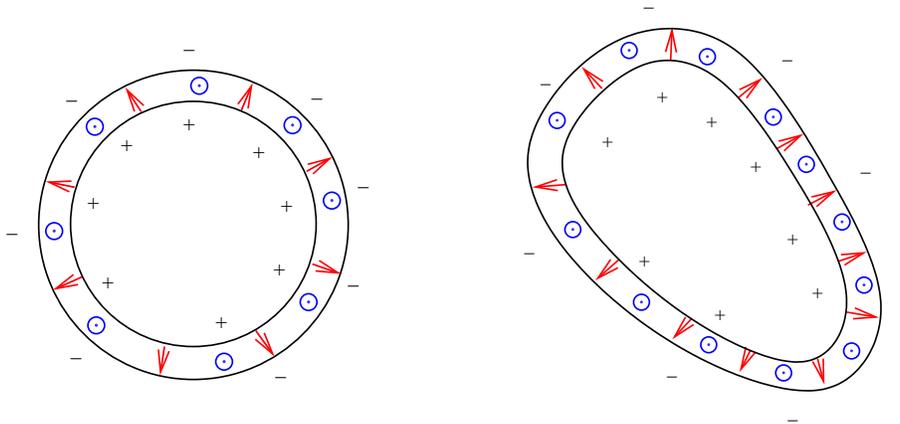}}
\caption{{\protect\small Moduli space of the $Q$ lump at large magnetic
flux. }}
\label{modulispace}
\end{figure}

Although, as was noted in Sect.~\ref{implqc}, the relation between these two
descriptions is not straightforward, one can think that the answer is
positive. From the tension formula (\ref{tension Qlump formula}) we see that
the same tension as for a ``round cylinder" is obtained for a generic
non-spherically symmetric $Q$ cylinder with the perimeter $2\pi \rho_*$. Any
closed curve with the fixed perimeter gives rise to an equally energetic $Q$
cylinder (see Fig.~\ref{modulispace}) as long as one can neglect terms
depending on the curvature of the cylinder walls. This can be viewed as a
remnant of the moduli space in the sigma-model description.

\begin{figure}[h!tb]
\epsfxsize=12cm \centerline{\epsfbox{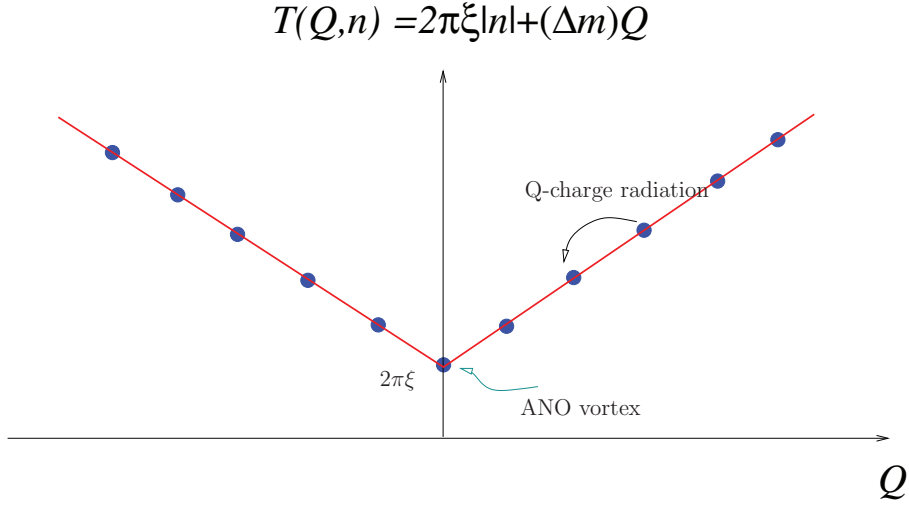}}
\caption{{\protect\small The $Q$ cylinder tension as function of $Q$ at
fixed $n$. The $Q$ cylinder is marginally stable with respect to radiation
of meson quanta with the $Q$ charge 1. }}
\label{qlumptension}
\end{figure}

Now let us discuss the issue of a possible $Q$ charge leakage from the
cylinder through the emission of elementary quanta ($Q$ stability). This
discussion runs in parallel to that of the wall $Q$ stability, Sect.~\ref%
{qwtea}. The $Q$ cylinder tension (see Fig.~\ref{qlumptension}) implies that
the $Q$ cylinder is marginally stable under the $Q$ charge radiation via
emission of mesons. Mesons, in the sigma model approximation, are the quanta
of the $u$ field. They have mass $\Delta m$ and carry $Q =1$. In the full
bulk theory the meson is a composite object composed of a quantum of the $q_2
$ field screened by a $\bar{q}_1$. From Table \ref{symmetries} we see that
this composite object has the vanishing electric charge and the U(1) charge $%
Q =1$ ($\frac12$ from the $q_2$ and $\frac12$ from $\bar{q}_1$).

Finally, let us discuss the fact that in the sigma model analysis the
frequency $\omega$ is fixed, while it seems to be arbitrary in the
worldvolume description.

We will show that in the bulk-theory description at the end of the day we
also get this feature, a specific value of the frequency. To this end we can
confront two expressions for the electric field obtained in different ways.
On the one hand, it is quite trivial that
\begin{equation*}
E=\frac{Q\,g^2}{2}\,\frac{1}{2\pi \rho}\,.
\end{equation*}
On the other hand, from the value of the gauge potential in Table~\ref%
{profilesQlump} we find
\begin{equation*}
E=\frac{2\omega}{d}\,.
\end{equation*}
Comparing these two expressions we get
\begin{equation}
\frac{\rho}{d}=\frac{Q\,g^2}{8 \pi \omega} \,,
\end{equation}
which in turn must be compared with Eq.~(\ref{randd}). The result is
consistent with (\ref{randd}) if $\omega =\Delta m$.

\begin{center}
\begin{table}[h!tb]
\begin{center}
\begin{tabular}{|c|c|c|}
\hline
\rule{0mm}{6mm} & $\rho =0$ & $\rho \rightarrow \infty $ \\ \hline
\rule{0mm}{6mm} $q_{1}\left( \rho ;\theta ,t\right) $ & $0$ & $\sqrt{\xi}
e^{-i\frac{\omega}{2} t}e^{in\theta }$ \\ \hline
\rule{0mm}{6mm} $q_{2}\left( \rho ;t\right) $ & $\sqrt{\xi} e^{i \frac{\omega%
}{2} t}$ & $0$ \\ \hline
\rule{0mm}{6mm} $\phi \left( \rho \right) $ & $-\sqrt{2} m_{1}$ & $-\sqrt{2}
m_{2}$ \\ \hline
\rule{0mm}{6mm} $A_{\theta }\left( \rho \right) $ & $0$ & $\frac{n}{\rho }$
\\ \hline
\rule{0mm}{6mm} $A_{t}\left( \rho \right) $ & $\omega $ & $-\omega $ \\
\hline
\end{tabular}%
\end{center}
\caption{{\protect\small Profile functions for the $Q$ string in the bulk
theory }}
\label{profilesQlump}
\end{table}
\end{center}

\subsection{Twisting the cylinder}

\label{ttc}

Now, we will make the next step in our program --- we will ``twist" the
magnetic flux lines. The easiest way to explain the twist is through the
worldvolume theory.

For an infinite flat wall the orientation of the magnetic field (which is in
one-to-one correspondence with $\vec\nabla\alpha$) is unimportant since we
can always align the $x$ axis with $\vec\nabla\alpha$. This ceases to be the
case when we compactify the $x$ direction. Let us stick to the convention
that the compactified direction is $x$. Typically, $\vec\nabla\alpha$ will
be misaligned. What is the physical picture behind this misalignment?

\begin{figure}[h!tb]
\epsfxsize=8cm \centerline{\epsfbox{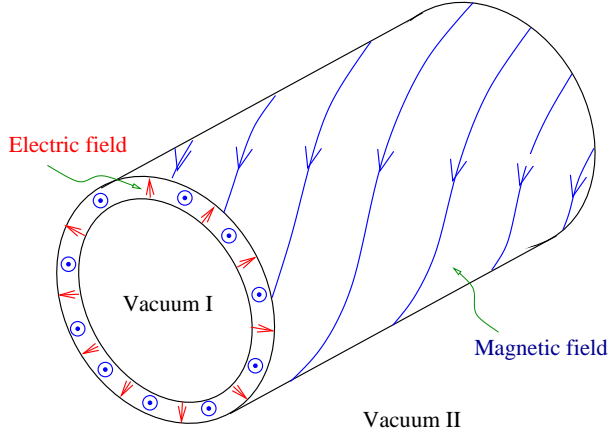}}
\caption{{\protect\small The flux line of the magnetic field inside the wall
at $a\neq 0,\,\,\, \tilde a\neq 0$. }}
\label{ctwi}
\end{figure}

It is quite obvious that compactifying the $x$ direction and
choosing
\begin{equation}
\alpha = a\, x + \tilde a\, y\,,\qquad a\neq 0,\,\,\, \tilde a\neq 0\,,
\end{equation}
we twist the magnetic field inside the wall with regards to the orientation
of the cylinder axis (Fig.~\ref{ctwi}). A similar situation was discussed in
the second paper in Ref.~\cite{ds}. The orientation of the magnetic field is
given by the vector
\begin{equation}
B_i \propto \varepsilon_{ij} a^j\,,\qquad a^j =\{ a,\tilde a\}\,.
\end{equation}

Note that in this geometry the coefficient $a$ is quantized while $\tilde a$
is not. At the next stage, when the cylinder is folded to make a torus, $%
\tilde a$ will be quantized too.

One must $Q$ charge the twisted cylinder to make it stable just in the same
vein as it was done in Sect.~\ref{cylinder} (see Fig.~~\ref{ctwi}).


\section{Twisted and $Q$ charged torus}

\label{tqct}

This section is central in the conceptual design of supersymmetric $Q$
solitons of arbitrary genus. Here we will explain how one can build a
genus-1 structure using the elements introduced above. To this end one glues
the twisted cylinder of Sect.~\ref{ttc} to produce a torus (see Fig.~\ref%
{potentialp}).
\begin{figure}[h!tb]
\begin{center}
\leavevmode \epsfxsize 8 cm \epsffile{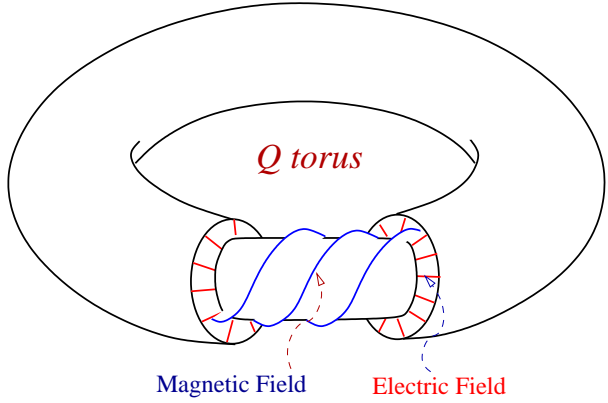}
\end{center}
\caption{{\protect\small Twisted torus.}}
\label{potentialp}
\end{figure}

If both sizes of the fundamental cell $L_x,\,\, L_y$ are large (compared to
the wall thickness) one can write
\begin{equation}
\alpha = 2\, \pi \, k \, \frac{x}{L_x} + 2\, \pi \, \tilde k \, \frac{y}{L_y}
\,,  \label{akxy}
\end{equation}
where the integers $k$ and $\tilde k$ quantize $a$ and $\tilde a$.
Correspondingly, the fluxes of the magnetic fields through both cycles of
the torus get quantized too. The third quantum number needed to get a stable
genus-1 soliton is its $Q$ charge.

\subsection{$Q$ torus in the full bulk theory}

\label{qtibt}

In the full bulk theory the $Q$ torus is a very complicate soliton; \textit{%
all} fields of the bosonic Lagrangian (\ref{n2sqed}) are exited. The only
symmetry that we can use is the cylindrical one.

Thus, we introduce cylindrical coordinates: $r ,z,\phi $. \ The profile
functions will have non-trivial dependence on both $r $ and $z$, and, in
some cases, also a phase rotation with regards to $\phi $ and $t$. The
boundary conditions are imposed on two circles (see Fig.~\ref%
{boundaryconditions}). One is the internal circle $z=0$ and $r =R$. The
other is the external circle composed by the axial line $r =0$ plus the
point at infinity. We summarize the profile functions and their boundary
conditions in Table \ref{profilesQtorus}.
\begin{figure}[h!tb]
\epsfxsize=7cm \centerline{\epsfbox{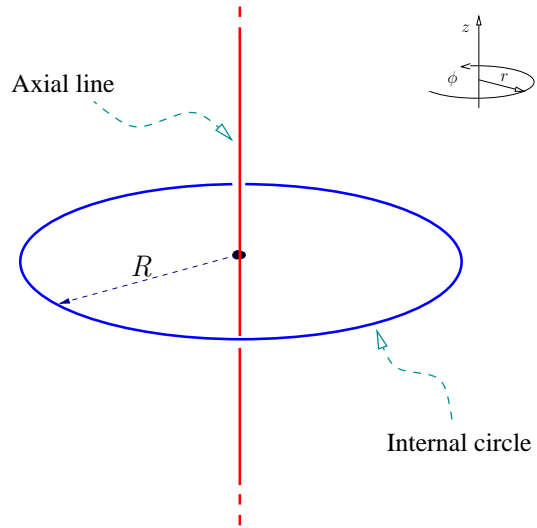}}
\caption{{\protect\small The boundary conditions for the $Q$ torus profile
functions (see Table \protect\ref{profilesQtorus}), are given on two
circles. The first one is the \textit{internal} circle (Vacuum $I$), the
other circle is the axial line plus the point at infinity (Vacuum $I\!I$).}}
\label{boundaryconditions}
\end{figure}

\begin{center}
\begin{table}[h!tb]
\begin{center}
\begin{tabular}{|c|c|c|}
\hline
\rule{0mm}{6mm} & Internal circle & Axial line \\ \hline
\rule{0mm}{6mm} $q_{1}\left( r ,z;\phi ,t\right) $ & $0$ & $\sqrt{\xi}e^{-i%
\frac{\omega}{2} t}f_{n}(z)$ \\ \hline
\rule{0mm}{6mm} $q_{2}\left( r ,z;\phi ,t\right) $ & $\sqrt{\xi}e^{i \frac{%
\omega}{2} t}e^{iN \phi }$ & $0$ \\ \hline
\rule{0mm}{6mm} $a \left( r ,z\right) $ & $-\sqrt{2} m_{1}$ & $-\sqrt{2}
m_{2}$ \\ \hline
\rule{0mm}{6mm} $A_{\phi }\left( r ,z\right) $ & $N/r $ & $0$ \\ \hline
\rule{0mm}{6mm} $A_{r }(r ,z)$ & $0$ & $0$ \\ \hline
\rule{0mm}{6mm} $A_{z}\left( r ,z\right) $ & $0$ & $-if_{n}^{\prime }\left(
z\right) /f_{n}\left( z\right) $ \\ \hline
\rule{0mm}{6mm} $A_{t}\left( r ,z\right) $ & $\omega $ & $-\omega $ \\ \hline
\end{tabular}%
\end{center}
\caption{{\protect\small Profile functions for the $Q$ torus in the bulk
theory.}}
\label{profilesQtorus}
\end{table}
\end{center}

A comment is in order here to explain not yet defined quantities in Table %
\ref{profilesQtorus} summarizing the boundary conditions. First of all, $R$
is the radius of the internal circle, also shown in Fig.~\ref%
{boundaryconditions}. This parameter is not fixed; it is to be determined by
energy minimization of the soliton. Another interesting quantity is the
function $f_n(z)$ that appears in the boundary condition on the axial line
for the scalar field $q_1$. This field must wind $n$ times around the vacuum
manifold $S^1$ when one passes from $-\infty$ to $+\infty$ on the axial
line. It is just a one-dimensional texture whose profile function $f_n(z)$
is not determined a priori and must be derived from energy minimization. The
$Q$ torus is thus quite a complicated solitonic object: it has another
``small'' soliton in its boundary conditions!

Needless to say, this object is quite resilient to direct solution even
using numerical methods since a set of entangled partial differential
equations is involved. The main purpose of this paper is to prove the
existence of an appropriate soliton and its stability. To this end we will
use the two approximations we exploited above to describe the $Q$ wall and $%
Q $ cylinder. Let us start from the bulk-theory description. In vein with
similar assumptions for the $Q$ wall and $Q$ cylinder, we will impose the
following constraint on the parameters of the $Q$ torus:
\begin{equation}
R\gg \rho \gg d\gg \delta _{\mathrm{edge}}\,.  \label{parli}
\end{equation}
As usual, $\delta _{\mathrm{edge}}$ is set to zero; thus the $Q$ torus
depends on three parameters: the total U(1) charge $q$, and two magnetic
fluxes $n$ and $N$. In the particular limit (\ref{parli}) we can write the
mass of the $Q$ torus as a function of three geometric parameters $R,\rho $
and $d$,
\begin{eqnarray}
M\left( R,\rho,d \right) &=& \underset{ \mathrm{Coulomb\ phase}}{\underset{}{
\frac{\pi ^{2}g^{2}\xi^{2}}{2}R \rho d }}+ \underset{\mathrm{Kinetic}\text{
\textrm{\ }}\ a }{\underset{}{\frac{8\pi ^{2}(\Delta m)^{2} }{g^{2}} \frac{%
R\rho}{d }}}  \notag \\[4mm]
& +&\underset{\mathrm{Magnetic\ flux}\text{ }n}{\underset{}{\frac{8\pi
^{2}n^{2}}{g^{2}}\frac{R}{\rho d }}}+\underset{\mathrm{Magnetic\ flux\ }N}{
\underset{}{\frac{8\pi ^{2}N^{2}}{g^{2}}\frac{\rho}{Rd }}}+\underset{
\mathrm{Electric\ field}}{\underset{}{\frac{q^{2}\, g^2}{32\pi ^{2}}\,\frac{
d }{R\rho }}}\,,  \label{loneq}
\end{eqnarray}
where we introduced the total charge (measured in integers)
\begin{equation}
2\pi \, R\, Q = q\,.
\end{equation}
The first and the second terms in Eq.~(\ref{loneq}) are obtained from those
in Eq.~(\ref{tension Qlump formula}) by multiplying by $2\pi R$. The third
and the fourth terms represent the third term in Eq.~(\ref{tension Qlump
formula}) which splits into two terms due to the fact that there are two
magnetic cycles now. The last term in Eq.~(\ref{loneq}) corresponds to the
last term in Eq.~(\ref{tension Qlump formula}).

To minimize this function we rewrite it in the following way:
\begin{eqnarray}
M(R,\rho,d )& =&\frac{\pi ^{2}g^{2}\xi^{2}R}{2}\rho d + \frac{8\pi ^{2}n^{2}R}{%
g^{2}} \frac{1}{\rho d }+  \notag \\[3mm]
& +&\frac{8\pi ^{2}(\Delta m)^{2}R}{g^{2}}\,\, \frac{\rho}{d }+\frac{8\pi
^{2}N^{2} }{g^{2}R}\frac{\rho }{d }+\frac{q^{2}\, g^{2}}{32\pi ^{2}R}\, \,%
\frac{d }{\rho}\,.  \label{massaa}
\end{eqnarray}
Now, the first line can be minimized with respect to $\rho d $ while the
second line can be minimized with respect to ${d }/{\rho}$, keeping $R$
fixed. In this way we arrive at
\begin{equation}
( \rho d)_* =\frac{4n}{g^{2}\xi}\,,\qquad \left(\frac{d}{\rho}\right)_* =%
\frac{16\pi ^{2}}{q\,g^2}\, \sqrt{(\Delta m)^{2}R^{2}+N^{2}}\,,
\label{lengthintermediate}
\end{equation}
where the asterisk marks the optimized $Q$-torus values. Inserting these
expressions back in Eq.~(\ref{massaa}) we get
\begin{equation}
M(R)=4\pi ^{2}\xi n R+\frac{q}{R}\, \sqrt{(\Delta m)^{2}R^{2}+N^{2}} \,.
\label{massa grande}
\end{equation}
Finally we can minimize $M(R)$ with respect to $R$ to obtain $M_{Q-\mathrm{%
torus}}$. For arbitrary $N$ the function will be rather complicated. In the
limiting case


\begin{equation}
(\Delta m) R\ll N
\end{equation}
the formula for $M_{Q-\mathrm{torus}}$ is simple, however. Indeed,
minimizing
\begin{equation}
M(R)=4\pi^{2}\xi n R+ \frac{q N}{R}  \label{4pixi}
\end{equation}
we find
\begin{equation}
R_{Q-\mathrm{torus}}=\frac{1}{2 \pi\sqrt{\xi}}\, \sqrt{\frac{qN}{n}}\,,
\qquad M_{Q-\mathrm{torus}}=4\pi \sqrt{\xi}\,\sqrt{qNn}\,.  \label{4pixip}
\end{equation}
Note that all the conditions in Eq.~(\ref{parli}) as well as $(\Delta m)
R\ll N$, can be satisfied under an appropriate choice of the charge $q$ and
the magnetic fluxes $n, N$. Equation (\ref{lengthintermediate}) implies
\begin{equation}
\rho_* =\frac{1}{2\pi\, \sqrt{\xi}}\sqrt{\frac{nq}{N}}\,, \qquad d_* = \frac{%
8\pi}{g^2\sqrt{\xi}}\sqrt{\frac{nN}{q}}\,.
\end{equation}
The conditions (\ref{parli}) now become
\begin{equation}
\frac{1}{2\pi\sqrt{\xi}}\sqrt{\frac{qN}{n}~} \gg\frac{1}{2\pi \sqrt{\xi}}\,%
\sqrt{\frac{nq}{N}}\gg \frac{8\pi}{g^2\sqrt{\xi}}\sqrt{\frac{nN}{q}}\gg\frac{%
1}{g\sqrt{\xi}}\,.
\end{equation}
Here $(g\sqrt\xi)^{-1}$ is the edge thickness, to be considered as the
smallest size in the problem at hand.

\vspace{2mm}

Multiplying by $\sqrt{\xi qNn}$ we obtain an equivalent constraint,
\begin{equation}
\frac{1}{{2}\pi}qN \gg \frac{1}{{2}\pi}nq \gg \frac{8\pi}{g^2} nN \gg \frac{1%
}{g} \sqrt{qNn} \,.  \label{condition one}
\end{equation}
The condition $(\Delta m) R\ll N$ implies
\begin{equation}
\sqrt{\frac{Nn}{q}}\gg\frac{\Delta m}{2\pi \sqrt{\xi}}\,.
\label{condition two}
\end{equation}
We remind that in the thin wall approximation $\Delta m/\sqrt\xi\gg 1$. It
is not difficult to see that Eqs. (\ref{condition one}) and (\ref{condition
two}) can always be satisfied with an appropriate choice of the integer
parameters
\begin{equation}
q\gg N\gg n\,,\qquad nN \gg q\,.
\end{equation}
The second of these two conditions is irrelevant since the constraint (\ref%
{condition two}) is stronger.

\vspace{2mm}

To conclude this section let us give our final result (in the thin edge
approximation). The $Q$ torus mass in this limit takes the form
\begin{equation}
M_{Q-\mathrm{torus}}=4\pi \sqrt{\xi}\,\sqrt{qNn} \left( 1+\mathcal{O} \left(
\frac{\Delta m }{\sqrt \xi}\sqrt{\frac{q}{Nn}}\right) \right)\,.
\label{massa approssimata}
\end{equation}

On the other hand, in the limit opposite to (\ref{condition two}),
\begin{equation*}
\sqrt{\frac{Nn}{q}}\ll\frac{\Delta m}{2\pi \sqrt{\xi}}\,,
\end{equation*}
when $q$ is very large, Eq.~(\ref{massa grande}) implies that
\begin{equation*}
\frac{q(\Delta m)}{n\xi}\gg R_*\gg \frac{N}{\Delta m}\,,
\end{equation*}
and
\begin{equation}
M_{Q-\mathrm{torus}}=q(\Delta m)\,.  \label{tueplu}
\end{equation}


\subsection{$Q$ torus in the sigma-model limit}

\label{qtsml}

Now let us look at the $Q$ torus soliton from a different perspective, that
of the sigma-model limit (i.e. $\Delta m \ll g\sqrt\xi$). It turns out that
in this limit $Q$ torus is related, in a way, to the Hopf Skyrmions
discovered in \cite{Faddeev:1996zj,Faddeev:1998eq} (for a recent work see
\cite{Bolognesi:2007ut}).

\begin{figure}[h!tb]
\epsfxsize=12cm \centerline{\epsfbox{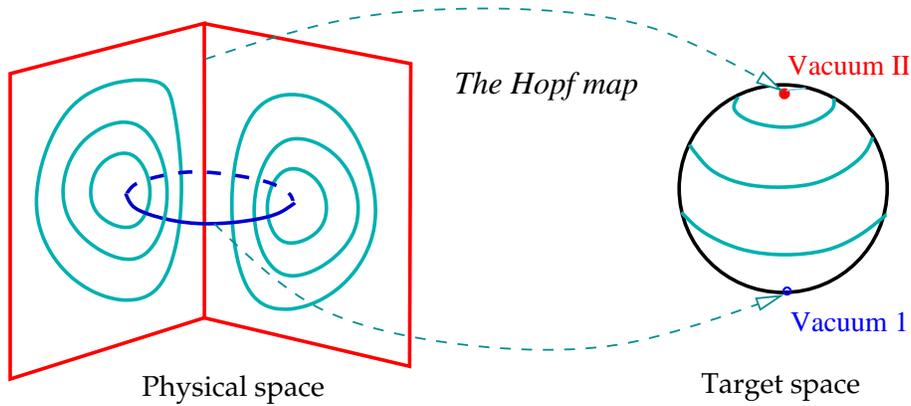}}
\caption{{\protect\small The Hopf map from the space $\mathbf{R}^2$ to the
target space $\mathbf{S}^2$. The axial line plus the points at infinity are
mapped onto the north pole (Vacuum $I\!I$). The internal circle is mapped
onto the south pole (Vacuum $I$). }}
\label{hopfmap}
\end{figure}

In the sigma-model approximation the $Q$ torus is a non-trivial element of
the Hopf homotopy group
\begin{equation}
\pi _{3}\left( \mathbf{S}^2\right) =\mathbf{Z} \, .
\end{equation}
An illustration that may help one to understand geometry of the Hopf map is
given in Fig.~\ref{hopfmap}.

Let us imagine the space $\mathbf{R}^3$ as a ``book.'' Every ``page'' of the
book is a semi-infinite plane attached to the axial line. The axial line
plus the points at infinity are mapped onto the north pole of the target
space $\mathbf{S}^2$. The pre-image of the south pole is a circle linked
with the axial line. Every semi-infinite plane is wrapped around the target
space.

The U(1) phase --- in Fig.~\ref{hopfmap} it is the rotation of $\mathbf{S}^2$
which keeps fixed the north and the south poles --- is twisted as the
semi-infinite plane is rotated around the axial line.

A comment is in order here to explain the stability of this Hopf Skyrmion.
In the theory described by the Lagrangian (\ref{lagrangiansigmamodel}), the
Derrick theorem forbids \cite{Derrick} the existence of particle-like
solitons. Even if the homotopy group creates nontrivial maps from the
coordinate space to the target space, the kinetic and the potential terms
for the scalar field cannot create a repulsive force necessary in order to
prevent the soliton from collapsing. A way out suggested in \cite%
{Faddeev:1996zj,Faddeev:1998eq} was adding higher derivative terms. To avoid
the Derrick collapse \emph{without introducing higher derivative terms} in
the Lagrangian we exploit a linear-in-time rotation of the U(1) phase which
stabilizes the soliton through a centrifugal force.


\subsection{Comment on the cycles' stability}

\label{comone}

A remark is in order here to better explain the stability of the cycles
relevant to the twisted torus. This remark will help us avoid a possible
source of confusion.

\begin{figure}[h!tb]
\epsfxsize=8.5cm \centerline{\epsfbox{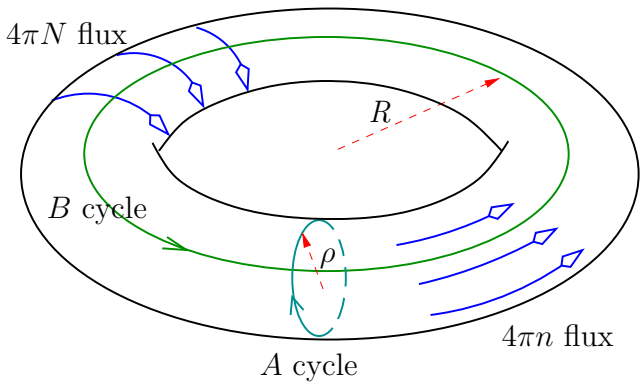}}
\caption{{\protect\small Two nontrivial homology cycles associated with the $%
Q$~torus: the $\protect A$ cycle with radius $\protect\rho$ and the $%
\protect B$ cycle with radius $R$. Note that the $\protect A$ cycle
is stabilized by the $4\protect\pi n$ magnetic flux that passes along the $%
\protect B$ cycle. And \emph{vice versa}, the $\protect B$ cycle is
stabilized by the $4\protect\pi N$ magnetic flux that goes along the $%
\protect A$ cycle.}} \label{stabilization}
\end{figure}

Consider the $Q$ torus in Fig.~\ref{stabilization}. The two
non-trivial cycles we have are denoted by $A$ and $B$ in this
figure. The $A$ cycle has radius $\rho$ while the $B$ cycle has
radius $R$. The magnetic fluxes are $4\pi n$ (the one that goes
parallel to the $B$ cycle) and $4\pi N$ (the one that goes parallel
to the $A$ cycle). A crucial point we want to stress is that the
magnetic flux stabilizes the cycle that crosses it rather than the
cycle which runs parallel to it.

For example, the flux $4\pi n$ is responsible for the stability of the $%
A $ cycle and \emph{vice versa}. In Sect.~\ref{qtibt} we obtained
quite simple formulae for the mass and dimensions of the $Q$ torus.
We can use these expressions to better understand the issue of
stability. First of all, let us note that $M_{Q-\mathrm{torus}}$ is
proportional to $\sqrt{qnN}$ which means that all three integer
parameters, $q$, $n$ and $N$, are crucial for the existence of the
$Q$ torus. If one of them vanishes the $Q$ torus disappears.

The radius $R$ is proportional to $\sqrt{{qN}/{n}}$ and we see that its
stability is guaranteed, in fact, by the electric field $q$ and the magnetic
flux $N$. The radius $\rho$ is proportional to $\sqrt{{qn}/{N}}$ implying
that its stability is guaranteed by $q$ and the magnetic flux $n$. The roles
of two magnetic fluxes are complementary.


\subsection{Interactions between $Q$ tori}

\label{isahg}


In this section we will discuss the issue of interaction between $Q$ tori.
This process is important if we want to establish a condition for the
existence of \textit{higher-genus} solitons. At the end of this section we
will be able to formulate this condition.

First of all let us visualize possible physical processes which are allowed
given conservation of fluxes and the U(1) charge. Let us consider the
physical process of \textit{merging} of $Q$ tori, that is two $Q$ tori
fusing to become a single one. Equivalently, if viewed in the opposite time
direction, this process describes a $Q$ torus decay into two smaller ones.
For simplicity we will concentrate only on the merging direction.

Two alternatives are possible depending on whether the cycle
involved is the $A$ cycle or the $B$ cycle (see
Fig.~\ref{stabilization} for our conventions regarding the cycle
names).

\begin{figure}[h!tb]
\begin{center}
\leavevmode \epsfxsize 12 cm \epsffile{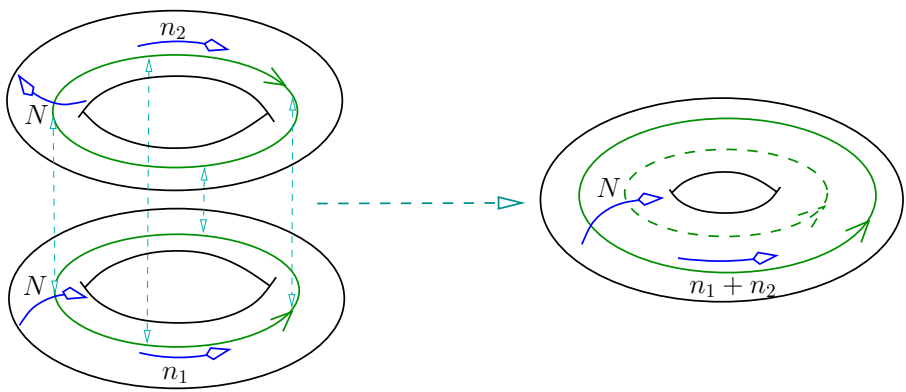}
\end{center}
\caption{{\protect\small A ``landing of two Flying Saucers'', or
merging of two $Q$ tori through their $\protect A$ cycles. This
process happens ``vertically." The two $Q$ tori must have the same
magnetic flux $N$; they
touch on their $\protect B$ cycles and then merge. We will refer to $%
\protect A$ merging or \textit{vertical} merging. Viewed in the
opposite direction it is an $\protect A$ or \textit{vertical}
decay.}} \label{verticaldecay}
\end{figure}

To begin with, let us deal with the $A$ cycle merging (it is
simpler). It is displayed in Fig.~\ref{verticaldecay} and
corresponds to the following process:
\begin{equation}
\left( n_{1},N,q_{1}\right) +\left( n_{2},N,q_{2}\right) \longrightarrow
\left( n_1 + n_2 ,N,q_1 +q_2\right) ~.  \label{merginguno}
\end{equation}
The $A$ process (it can be called ``vertical merging" as is clearly
seen from Fig.~\ref{verticaldecay}), is possible only if two fusing
$Q$ tori share the same $N$ flux.

\begin{figure}[h!tb]
\begin{center}
\leavevmode \epsfxsize 12 cm \epsffile{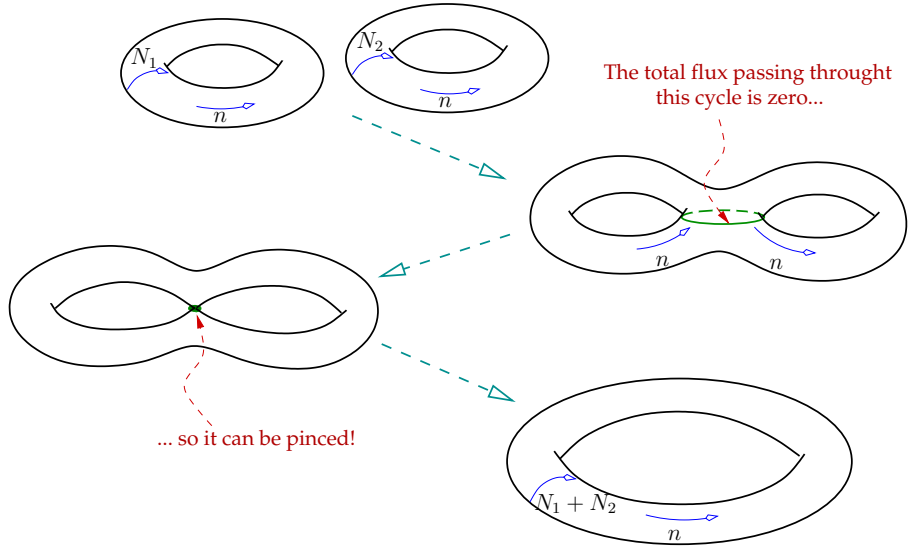}
\end{center}
\caption{{\protect\small A ``crush of two Flying Saucers'' or
merging of two $Q$ tori through their $\protect B $ cycles. This
process happens horizontally. The two merging $Q$ tori must have the
same magnetic flux $n$. They touches and form a genus-$2$ $Q$
soliton. In fact, this is not a genuinely genus-$2$ soliton: the
middle handle has vanishing total magnetic flux passing trough it.
Therefore, this handle is unstable and can be
pinched away. What is left is a $Q$ torus of genus $1$. We can call it $%
\protect B$ or \textit{horizontal} merging. In the opposite
direction it is a $\protect B$ or \textit{horizontal} decay.} }
\label{horizontaldecay}
\end{figure}

The second process which we call $B$ merging (or horizontal merging)
corresponds to merging of two $B$ cycles of two merging $Q$ tori.
The corresponding reaction is
\begin{equation}
\left( n,N_1,q_{1}\right) +\left( n,N_2,q_{2}\right) \longrightarrow \left(
n,N_1 +N_2,q_1 +q_2\right) \,,  \label{mergingdue}
\end{equation}
see Fig.~\ref{horizontaldecay}. As is perfectly clear from Fig.~\ref%
{horizontaldecay} this process is more convoluted than the vertical one. It
is possible only if two merging $Q$ tori share the same $n$ flux.

The conditions imposed by the flux and charge conservation is not the end of
the story. We must discuss the energy balance. This is a more complicated
issue since it involves knowledge of the mass function $M_{Q-\mathrm{torus}}$
which, so far, we have only in a certain limit. But we can at least
speculate on the marginal stability.

For generic solitons which depend on a single integer, say, $n$, the mass is
said to be marginally stable if it is a linear function of $n$. There is
only one process of merging of such solitons, namely, an $n_1$ soliton adds
to an $n_2$ soliton to form a compound $n_1 +n_2$ soliton. Marginal
stability with respect to this process constrains the mass function to be a
linear function of $n$.

In the case under discussion, the $Q$ torus, the soliton mass depends on
three integers, and two processes are allowed. The marginal stability
condition does constrain the mass function but still leaves a lot of freedom.

Note that minimization of (\ref{massa grande}) gives a function which is
marginally stable (see e.g. (\ref{massa approssimata})). We can easily prove
this assertion by considering only the relevant parts of the formula (\ref%
{massa grande}),
\begin{equation*}
M(R)=nR+\frac{q}{R}\sqrt{R^{2}+N^{2}}\,,
\end{equation*}
where for simplicity we set all irrelevant parameters to 1. Under the
``vertical" rescaling
\begin{equation*}
\left( n,N,q\right) \rightarrow \left( an,N,aq\right)
\end{equation*}
we have to make the simultaneous rescaling
\begin{equation*}
M\rightarrow aM\,\,\,\mathrm{and}\,\,\, R\rightarrow R\,.
\end{equation*}
Under the ``horizontal" rescaling
\begin{equation*}
\left( n,N,q\right) \rightarrow \left( n,bN,bq\right)
\end{equation*}
we have to make the simultaneous rescaling
\begin{equation*}
M\rightarrow bM\,\,\,\mathrm{and}\,\,\, R\rightarrow bR\,.
\end{equation*}
The geometric meaning of these rescalings is represented in Fig.~\ref%
{rescaling}.
\begin{figure}[h!tb]
\begin{center}
\leavevmode \epsfxsize 11 cm \epsffile{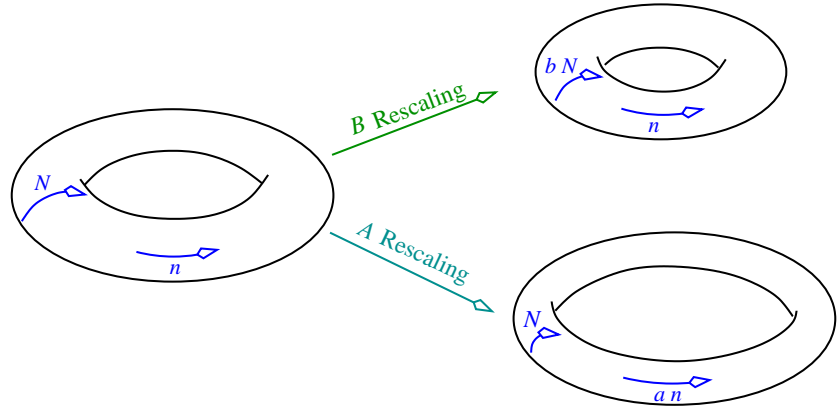}
\end{center}
\caption{{\protect\small The $\protect A$ and the $\protect B$
rescalings. }} \label{rescaling}
\end{figure}


\subsection{Higher genera}

\label{hige}

\begin{figure}[h!tb]
\begin{center}
\leavevmode \epsfxsize 9 cm \epsffile{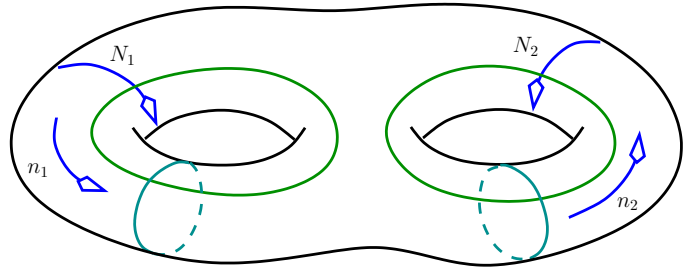}
\end{center}
\caption{{\protect\small A $Q$ soliton of genus 2.}}
\label{genustwo}
\end{figure}

Now we can present a condition for a genus-2 $Q$ soliton to be a \textit{%
genuine} genus-2 object and not something that could decay into a genus-1 $Q$
torus. In Fig.~\ref{genustwo} we have depicted a $Q$ soliton with its
nontrivial cycles and fluxes. The soliton now depends on five charges: the
electric charge $q$, the magnetic fluxes $n_1$ and $n_2$ which stabilize the
$A$ cycles and the magnetic fluxes $N_1$ and $N_2$ which stabilize the $%
B$ cycles. The soliton is genuinely genus-2 if the two processes (\ref%
{merginguno}) and (\ref{mergingdue}) are forbidden, namely
\begin{equation}
n_1 \neq n_2 \ , \quad \mathrm{and} \quad N_1 \neq N_2 \, .  \label{gen2}
\end{equation}

For higher genera the condition is more complicated, not just a
straightforward generalization of (\ref{gen2}). For example, take a genus-3
soliton with the fluxes $(1,1)$, $(2,2)$ and $(3,3)$. At first sight, it
might seem a genuine genus-3 object since none of the integers $n$ are equal
and so are $N$'s (they are unequal too). This implies that it cannot be
reduced to lower-genus solitons through only one transition.

However, we must consider a possibility of multiple combined transitions.
For instance, the last handle can split with by vertical decay into $(1,3)$
plus $(2,3)$. Then we will have four handles $(1,1)$, $(1,3)$, $(2,3)$ and $%
(2,2)$. The first two can now horizontally merge into $(1,4)$ while the last
two can horizontally merge into $(2,5)$. An apparent genus-3 soliton thus
turned out to be a genus-2 soliton. The mathematical condition for genuine
higher-genera solitons taking into account these multiple transition of
combined decays/mergings is yet to be found.

\subsection{$Q$ charge radiation revisited}

\label{qchar}

The leakage of the U(1) charge through radiation of elementary $Q$
charged quanta was discussed more than once above. Here we would
like to return to this question in connection with $Q$ stability of
the $Q$ torus. This is a dynamical question and its answer crucially
depends on the still unknown function $M(n,N,q)$. To find out
whether or not a generic $Q$ torus $(n,N,q)$ is stable under the
decay into a $Q$ torus $(n,N,q-1)$ plus a ``meson" we have to
confront the mass difference $M(n,N,q)-M(n,N,q-1)$ with the meson
mass.

To analyze a concrete example let us take the mass formula (\ref{massa
grande}) which in the case $\Delta m R\ll N$ leads to Eq.~(\ref{4pixip}) for
$M_{Q-\mathrm{torus}}$. From the square root $q$ dependence of $M_{Q-\mathrm{%
torus}}$ we deduce that there is a critical value
\begin{equation}
q_* \sim \sqrt[3]{\frac{\xi nN}{(\Delta m)^2}}
\end{equation}
below which the $Q$ torus is instable under the $Q$ charge radiation. Above $%
q_*$ the $Q$ torus is stable. This is not the end of the story, however.

\begin{figure}[h!tb]
\epsfxsize=12cm \centerline{\epsfbox{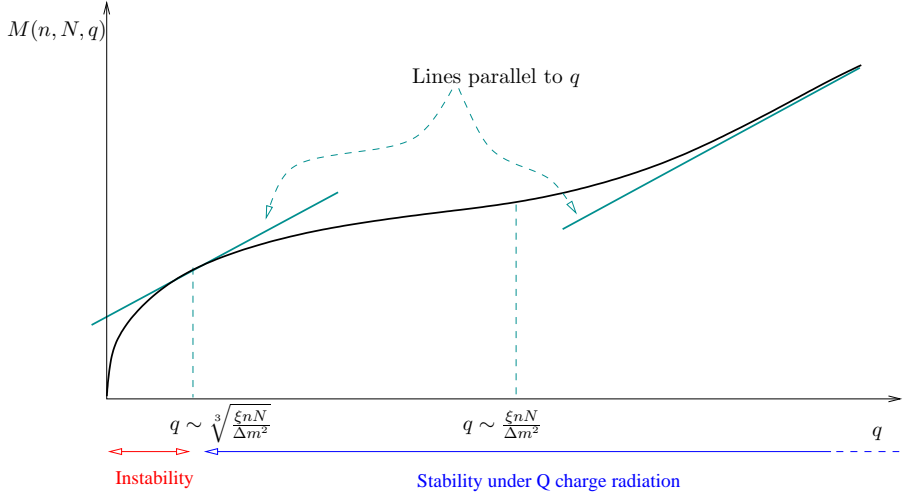}}
\caption{{\protect\small A schematic representation of $M_{Q-\mathrm{torus}}$
vs. $q$ at fixed $n$ and $N$. }}
\label{qtorustension}
\end{figure}

To analyze the problem of $Q$ stability for arbitrary large values of $q $
we have to return to the mass formula (\ref{massa grande}) since the
approximation $(\Delta m) R\ll N$ will certainly break down at
\begin{equation}
q \sim \frac{\xi nN}{(\Delta m)^2}\,.
\end{equation}
At $q \to \infty$ we can neglect the first term in Eq.~(\ref{massa grande})
and directly minimize the second term. The minimum $M = q\left(\Delta
m\right)$ is achieved at very large $R$ ($R\to\infty$). This mass function,
see Eq.~(\ref{tueplu}), is (asymptotically) tangential to the line of
marginal stability. In Fig.~\ref{qtorustension} we sketch a qualitative
behavior of the $Q$-torus mass as a function of $q$.

\section{The Dirac--Born--Infeld Action}

\label{tdbia}

In Sects. \ref{fieldexited}, \ref{cylinder}, and \ref{tqct} we used
the worldvolume theory framework to consider $Q$ walls, $Q$
cylinders and $Q$ tori. The effective $(2+1)$-dimensional action (\ref%
{21theory}) describes low-energy dynamics of two massless moduli on the
worldvolume of the domain wall, namely the translational fluctuation $z_0$
and the U(1) phase $\alpha$. Alternatively, one can deal with $(2+1)$%
-dimensional ${\mathcal{N}}=2\,$ \, SQED (\ref{pure}). The expressions for
the action we used in these sections were derived in \cite{Shifman2002};
they are valid to the lowest order in derivatives (quadratic in
derivatives). Quartic and higher order derivative terms are not included in
Eq.~(\ref{pure}).

Needless to say, the full worldvolume action contains higher-order
derivative terms. For weak magnetic and electric fields, in the low-energy
limit, higher order-terms can be neglected. Then we have a free field theory
(\ref{21theory}) containing only the kinetic terms. However, higher-order
terms are crucial (even in the low-energy limit) if we want to describe
solutions with nontrivial geometry, such as the $Q$ string or $Q$ torus.
After minimization, at the optimal values of $R$ and $d$, the magnetic and
electric fields stabilizing the solitons need not be small.

The free theory (\ref{21theory}) can describe weakly charged $Q$ walls, as
we saw in Sects.~\ref{cmfitw} and \ref{mean}, with crossed electric and
magnetic fields, but it cannot fully reflect nontrivial geometry, say, of
the $Q$ torus.

In order to obtain the effective worldvolume action including higher-order
derivative terms we can rely on two very well-known principles: geometry and
supersymmetry. Geometry tells us that the action on the brane must contain a
term proportional to the volume of the domain wall. If we call $%
x^{\alpha}=(t,x,y)$ the coordinates on the wall worldvolume and $%
X^{\mu}=(T,X,Y,Z)$ the coordinates of space-time,\footnote{%
In this section we adopt the metric conventions $\eta_{\gamma\beta}=\left(
-1,1,1\right)$ for three-dimensional theory and $\eta_{\mu\nu}=\left(
-1,1,1,1\right)$ for four-dimensional theory.} the action must contain the
term
\begin{equation}
-\int d^3 x \sqrt{-\det{g_{\gamma\beta}}} \,,
\end{equation}
where $g_{\alpha\beta}$ is the pull-back of the space-time metric on the
worldvolume.

Fixing the gauge of the local coordinates we can write the metric as
\begin{equation}
g_{\alpha\beta}=\eta_{\gamma\beta}+\partial_{\gamma} z \partial_{\beta} z \,,
\end{equation}
where $z$ is the translational modulus fluctuation (for brevity we omit the
subscript 0). As we know, $z$ is the superpartner of the U(1) modulus $%
\alpha $; therefore, requiring supersymmetry we arrive at
\begin{equation}
S=-T_{\mathrm{w}}\, \int d^3 x \sqrt{-\det\left[ {g_{\gamma\beta}+ \frac{1}{%
(\Delta m )^2}\, \partial_{\gamma} \alpha\, \partial_{\beta} \alpha}\right]}
\,.  \label{galp}
\end{equation}
Dualizing the U(1) phase according to Eq.~(\ref{21gaugenorm}) we arrive at
the Dirac--Born--Infeld (DBI) action
\begin{equation}  \label{DBIaction}
S=-T_{\mathrm{w}}\, \int d^3 x \sqrt{-\det\left[ g_{\alpha\beta}+ \frac{1}{%
2\pi\,\xi}\,F_{\alpha\beta}\right] }\,.
\end{equation}
If we expand the DBI action in the fields we get to the leading order
\begin{equation}
S=\int d^3 x \left[-T_{\mathrm{w}}- \frac{T_{\mathrm{w}}}{2}%
\partial_{\gamma}z\, \partial^{\gamma}z -\frac{1}{4 \, e^2 }\,
F_{\gamma\beta}F^{\gamma\beta}+\dots\right],
\end{equation}
which exactly coincides with Eq.~(\ref{21theory}).

It is instructive to note that the coefficients in Eq.~(\ref{DBIaction})
automatically match standard expressions of string theory. The standard
relative coefficient between $F_{\gamma\beta}$ and $g_{\gamma\beta}$ is $%
2\pi\alpha^\prime$ where $\alpha^\prime$ is the slope of the Regge
trajectory expressible in terms of the string tension as
\begin{equation*}
T_{\mathrm{string}} =\frac{1}{2\pi\alpha^\prime}\,.
\end{equation*}
(For an introductory presentation see e.g. \cite{joh}.) Therefore,
approaching the DBI action from this side we we would expect the relative
coefficient $1/T_{\mathrm{string}}$. That's exactly what we have in  Eq.~(%
\ref{DBIaction}) since the tension of the ANO string in the problem at hand
is  $2\pi\xi$.

At the moment it is not clear in which sense the DBI action is complete and
adequate in the problem at hand. Certainly, it \emph{ceases to present
adequate description} at a certain energy scale where heavy degrees of
freedom, either on the brane or in the bulk, must be taken into account.
This assertion is true both in string theory, where we deal with the D$2$
brane of Type IIA superstring, and in our model where the brane is a domain
wall of ${\mathcal{N}}=2\,$\, SQED.

Another important point to be noted here is that the DBI action has a
maximal value of the electric field. While the physical meaning of this is
well understood in string theory, at the moment it is not quite clear in the
field-theoretic context. We take it as a signal that the worldvolume
description in terms of the moduli fields fails and other degrees of freedom
become important. It could be related to the creation of a ``vortex handle''
on the worldvolume. This effect, essentially classical from the bulk point
of view, is a quantum polarization of the vacuum from the $(2+1)$%
-dimensional standpoint.

The DBI action is known to have interesting solutions where the crossed
electric and magnetic fields affect global geometry of the brane. One
example is the \textit{supertube} solution \cite{Mateos:2001qs}. The
supertube is a D$2$ brane with cylindrical geometry. It is a bound state of F%
$1$ strings and D$0$ branes which in the large-flux limit becomes well
described by the cylindrical D$2$-brane. In the field-theoretic context it
corresponds to the $Q$ string \cite{GPTT} (see our discussion in Sect.~\ref%
{implqc}). Feld-theoretic implementations of the supertube solution in other
(albeit related) models are discussed in \cite{Lee}.

Now we will show that the DBI action admits a solution with toric geometry
which corresponds to the $Q$ torus we discussed at length in the previous
sections.


\subsection{Construction of the $Q$ torus from the DBI action}

\label{constrc}

Let us parametrize the worldvolume coordinates as $x^{\alpha}=\left(
t,\theta,\phi\right) $ where $\theta,\phi$ are two angles defined on the
interval $[0,\, 2\pi ]$. The space-time is parametrized by cylindrical
coordinates $X^{\mu }=\left( T,Z,R,\Phi\right) $.


We fix the worldvolume reparametrization invariance in the following way:
\begin{eqnarray}
&& T = t\,,\qquad\Phi=\phi\,; \\[2mm]
&& R\left( \theta+2\pi\right) = R\left( \theta\right) \,; \\[2mm]
&& Z\left( \theta+2\pi\right) = Z\left( \theta\right) \,.  \label{wvrep}
\end{eqnarray}
The only reparametrization freedom left is $\theta\rightarrow f\left(
\theta\right) $. The induced metric on the worldvolume is
\begin{equation}
\partial_{\gamma }X^{\mu}\partial_{\beta}X_{\mu}=\left(
\begin{array}{ccc}
-1 & 0 & 0 \\
0 & R^{\prime2}+Z^{\prime2} & 0 \\
0 & 0 & R^{2}%
\end{array}
\right) ~.
\end{equation}
The field strength is
\begin{equation}
F_{\gamma \beta}=\left(
\begin{array}{ccc}
0 & E_{\theta} & E_{\phi} \\
-E_{\theta} & 0 & B \\
-E_{\phi} & -B & 0%
\end{array}
\right).
\end{equation}
Remember that now we deal with the \emph{dual} formulation of the bulk
theory: the magnetic field corresponds to the electric field in bulk and
\emph{vice versa}. With our coordinate choice we can now rewrite the DBI
action as follows:
\begin{equation}
S =- \int\,d^3x \sqrt{R^{2}\left( R^{\prime2}+Z^{\prime2}\right)
-E_{\phi}^{2}\left( R^{\prime2}+Z^{\prime2}\right) -
R^{2}E_{\theta}^{2}+B^{2}}\,.
\end{equation}
For simplicity we use the DBI action (\ref{DBIaction}) setting the
parameters
\begin{equation}
T_{\mathrm{w}}=1\,\,\,\, \mathrm{and} \,\,\,\,\frac{1}{2\pi\xi} =1\, .
\end{equation}
In the end we will reintroduce the couplings to match our results with the
ones of the previous sections. From now on, to simplify the notation, we
introduce
\begin{equation}
l=\sqrt{R^{\prime 2}+Z^{\prime2}}\,.
\end{equation}

Next, we want to pass from the Lagrangian to Hamiltonian. To this end we
must define the canonic momentum conjugated to the electric field,
\begin{equation}
\Pi^{\theta}=\frac{\delta\mathcal{L}}{\delta E_{\theta}}~,\qquad\Pi^{\phi }=
\frac{\delta\mathcal{L}}{\delta E_{\phi}}~,
\end{equation}
which leads to the following expressions:
\begin{eqnarray}
E_{\theta }& =&\frac{\Pi ^{\theta }}{R}\sqrt{\frac{l^{4}R^{2}+l^{2}B^{2}}{
l^{2}R^{2}+\Pi ^{\phi 2}R^{2}+l^{2}\Pi ^{\theta 2}}} \,, \\[5mm]
E_{\phi }& =&\frac{\Pi ^{\phi }r}{l}\sqrt{\frac{B^{2}+l^{2}R^{2}}{\Pi ^{\phi
2}R^{2}+l^{2}\Pi ^{\theta 2}+l^{2}R^{2}}}\,.
\end{eqnarray}
As a result, the Hamiltonian takes the form
\begin{eqnarray}
\mathcal{H}& =&\Pi ^{\theta }E_{\theta }+\Pi ^{\phi }E_{\phi }-\mathcal{L}
\notag \\[3mm]
& =&\frac{1}{lR}\sqrt{\left( B^{2}+R^{2}l^{2}\right) \,\,\left( \Pi ^{\theta
2}l^{2}+R^{2}l^{2}+\Pi ^{\phi 2}R^{2}\right) }\,.
\end{eqnarray}
The mass of the solitonic object to be determined is
\begin{equation}
M=\oint d\theta \oint d\phi ~\mathcal{H}\,,
\end{equation}
with the following conserved charges
\begin{equation}  \label{integers}
n=\oint d\theta ~\Pi ^{\phi }~,\quad N=\oint d\phi ~\Pi ^{\theta }~,\quad
2\pi q=\oint d\theta \oint d\phi ~B\,.
\end{equation}
In the end of this Section we will explain why this is the correct way to
relate the charges obtained from the $2+1$ effective action with the ones
obtained in the bulk. But for the moment let us stress that the charges $%
q,n,N$ in (\ref{integers}) are exactly the ones used in Sect.~\ref{qtibt}.

To understand some properties of this toric solution, we have to work in
some limit where we are able to solve the problem. Of particular interest is
the adiabatic limit in which the radius of the torus $R$ is much bigger than
the radius of the tube $\rho$,
\begin{equation}
l= \sqrt{R^{\prime 2}+Z^{\prime 2}}=\rho\ll R\,.
\end{equation}
In this case we have
\begin{equation}
\Pi ^{\phi}=\frac{n}{2\pi}, \,\,\, \Pi ^{\theta }=\frac{N}{2\pi},\,\,\,
\mathrm{and} \,\,\, B=\frac{q}{2\pi}\,,  \label{bltue}
\end{equation}
all independent of $\theta $. The mass of the soliton is
\begin{equation}  \label{masstominimize}
M\left( \rho,R\right) =\frac{1}{\rho R}\sqrt{\left( q^{2}+4 \pi^2
R^{2}\rho^{2}\right) \,\,\left( N^{2}\rho^{2}+4\pi^2
R^{2}\rho^{2}+n^{2}R^{2}\right) }\,.
\end{equation}
The above expression must be minimized with respect to $\rho$ and $R$. It is
clear that such a minimum exists.

Before we pass to minimizing this expression, we can immediately infer some
interesting property. First of all, the mass coming from the minimization of
(\ref{masstominimize}) is marginally stable with respect to the vertical and
horizontal decays. Indeed, under the ``vertical" rescaling
\begin{equation*}
\left( n,N,q\right) \rightarrow \left( an,N,aq\right)
\end{equation*}
we have to make a simultaneous rescaling of $M$, $\rho$ and $R$,
\begin{equation*}
M\rightarrow aM,\,\,\, R\rightarrow R,\,\,\,\mathrm{and } \,\,\,
\rho\rightarrow a\rho\,;
\end{equation*}
while under the ``horizontal" rescaling
\begin{equation*}
\left( n,N,q\right) \rightarrow \left( n,bN,bq\right)
\end{equation*}
we have to make a rescaling  of the following form:
\begin{equation*}
M\rightarrow bM,\,\,\, R\rightarrow bR,\,\,\,\mathrm{and } \,\,\,
\rho\rightarrow \rho\,.
\end{equation*}

Now let us try to minimize the mass function (\ref{masstominimize}). First
of all, calculating the derivatives with respect to $R$ and $\rho$, we can
get two relations,
\begin{equation}
R=\frac{1}{\sqrt{2\pi}}\frac{\sqrt{qN}}{\sqrt[4]{4\pi^2 \rho^2+n^2}} \, ,
\qquad \rho=\frac{1}{\sqrt{2\pi}}\frac{\sqrt{qn}}{\sqrt[4]{4\pi^2 R^2+N^2}}
\, .
\end{equation}
Next, performing the substitutions $R=a\sqrt{\frac{qN}{n}}$ and $\rho=b\sqrt{%
\frac{qn}{N}}$ we obtain two equations for $a$ and $b$,
\begin{equation}
a=\frac{1}{\sqrt{2\pi}\sqrt[4]{4\pi^2 b^2 \frac{q}{nN}+1}} \, , \qquad b=%
\frac{1}{\sqrt{2\pi}\sqrt[4]{4\pi^2 a^2 \frac{q}{nN}+1}}\,.
\end{equation}
Again, as was observed in Sect.~\ref{qtibt}, the solution is governed by the
parameter ${q}/(nN)$. Thus we have to distinguish two regimes.

\vspace{2mm}

(i) $q \ll nN$:

\vspace{1mm}

\noindent In this regime $a=b=\frac{1}{\sqrt{2\pi}}$. The radia are
\begin{equation}
R=\frac{1}{\sqrt{2\pi}}\sqrt{\frac{qN}{n}} \ , \qquad \rho= \frac{1}{\sqrt{%
2\pi}}\sqrt{\frac{qn}{N}}\, .
\end{equation}
while the soliton mass is
\begin{equation}  \label{uno}
M(n,N,q)=2\sqrt{2\pi}\sqrt{qnN}\,.
\end{equation}

\vspace{2mm}

(ii) $q \gg nN$:

\vspace{1mm}

\noindent In this regime
\begin{equation*}
a=b=\frac{1}{(2\pi)^{7/6}}\sqrt[6]{\frac{nN}{q}}\,.
\end{equation*}
The radia are
\begin{equation}
R=\sqrt[3]{\frac{1}{4\pi^2}\frac{N^2 q}{n}} \ , \qquad \rho=\sqrt[3]{\frac{1%
}{4\pi^2}\frac{n^2 q}{N}} \, ,
\end{equation}
while the soliton mass is
\begin{equation}  \label{due}
M(n,N,q)=2\pi q\,.
\end{equation}

\vspace{1mm}

To compare the above results with those of Sect.~\ref{qtibt} we must
reintroduce back the couplings in the action. According to (\ref{DBIaction})
all lengths must be rescaled by the factor $\sqrt[3]{T_{\mathrm{w}}^{-1}}$
while the field strength must be rescaled by the factor $\frac{1}{2\pi \xi}$%
. Skipping all passages of the Hamiltonian procedure, we find the new
version of the mass formula (\ref{masstominimize}) which reads
\begin{eqnarray}
M\left( \rho,R\right) &=&\frac{\Delta m \xi}{\rho R}\sqrt{ \frac{1}{4 \pi^2
\xi^2}q^{2}+4 \pi^2 R^{2}\rho^{2}}  \notag \\[4mm]
& \times & \sqrt{ \frac{4 \pi^2}{ \Delta m^2}N^{2}\rho^{2}+4\pi^2
R^{2}\rho^{2}+\frac{4 \pi^2}{ \Delta m^2}n^{2}R^{2}}\,.
\label{newmasstominimize}
\end{eqnarray}

\vspace{1mm} \noindent The final expressions for the limiting cases we are
interested in are

\vspace{2mm}

(i) $q \ll nN$:

\vspace{1mm}

\noindent The soliton mass is
\begin{equation}  \label{unop}
M(n,N,q)=4\pi\sqrt{\xi}\sqrt{qnN}\,.
\end{equation}

(ii) $q \gg nN$:

\vspace{1mm}

\noindent In this regime the soliton mass is

\begin{equation}  \label{duep}
M(n,N,q)=(\Delta m) q\,.
\end{equation}

\vspace{1mm} \noindent We note with satisfaction that the mass formulae we
have just obtained from the analysis of the DBI action perfectly match our
previous results in Sect.~\ref{qtibt}.

Before concluding this section we have to return to Eqs.~(\ref{integers})
and (\ref{bltue}) and show that this is indeed the correct way to match the
brane and the bulk parameters.

Let us start from the magnetic field. Consider the $Q$ wall studied in Sect.~%
\ref{mean} with a time-dependent rotation of the phase $\alpha$ given by (%
\ref{lint}). The $(2+1)$-dimensional magnetic field generated by this
rotation is
\begin{equation*}
F^{(2+1)}_{12}=2\pi\, \frac{\xi\, \omega}{\Delta m}\,.
\end{equation*}
From the bulk standpoint we have an electric field inside the capacitor $%
E_{z}=\sigma$. From Eq.~(\ref{segtpp}) we have the relation between $\omega$%
, the phase velocity, and $\sigma$, the charge density on the capacitor. The
electric field is thus
\begin{equation*}
E_z = \frac{g^2}{2}\frac{\omega \xi}{\Delta m}\,.
\end{equation*}
This justifies the relation (\ref{integers}) for the magnetic field on the
brane.

To justify the choice for the electric field conjugates quoted in Eq.~(\ref%
{integers}), we use the ANO vortex ending on the domain wall. From the point
of view of the bulk, the vortex is a flux tube carrying $4\pi n$ units of
the magnetic flux. From the brane standpoint the vortex is an electrically
charged particle. The electric flux must be measured with the electric
displacement $\Pi=\frac{\delta\mathcal{L}}{\delta E}$ which, for small brane
curvatures, is
\begin{equation*}
\Pi=\frac{E}{e^2}=\frac{E\Delta m}{4\pi^2 \xi}\,.
\end{equation*}
Using Eq.~(\ref{21gaugenorm}) we can thus express the electric charge as the
circulation of the phase $\alpha$ and we finally get
\begin{equation*}
\int \Pi = n\,,
\end{equation*}
which confirms the first two equation in (\ref{integers}).




\section{Conclusions and open questions}

\label{conoq}

Supersymmetric gauge theories are full of surprises and they have now
revealed another one: the occurrence of the $Q$ torus and other $Q$ solitons
of arbitrary genus! All these solitons are built from a folded domain wall
which supports an ${\mathcal{N}}=2\,$ Abelian gauge theory on its
worldvolume. The domain wall is folded around closed and oriented
two-dimensional surfaces. The stabilization of various nontrivial cycles, or
handles, is a consequence of magnetic fluxes inside the wall and
perpendicular electric fields.

The $Q$ charge obtained by making a time dependent rotation of the U(1)
modulus is crucial for stabilization. After the $Q$-charging procedure the
domain wall becomes similar to a capacitor. There are two edges with charge
respectively $+q$ and $-q$. Between the two edges there is a Coulomb phase
(an instable but stationary point of the potential). At the exterior of the
two edges there are respectively vacuum $I$ and vacuum $I\!I$. In the
interior Coulomb phase there are magnetic fluxes that flow along some
nontrivial cycles and cross other nontrivial cycles. For instance, the
magnetic flux going around the $\alpha$ cycle passes through the $\beta$
cycle and \emph{vice versa}. The stabilization of a given cycle occurs
through a combination of the magnetic flux that passes through it and the
orthogonal electric field created by the $Q$ charging procedure. Both
ingredients are equally crucial.

Now it is instructive to comment on the relation between the sigma models,
where the toroidal-twisted solitons are stabilized by the Hopf quantum
number, and other approaches to $Q$ tori. Clearly, there is a mismatch that
must be explained. In the sigma-model limit the soliton is labeled by two
numbers, the Hopf charge which we call $k$ (it is the integer invariant of $%
\pi_3(\mathbf{(}S^2)$), and the $q$ charge. In other approaches we have
three numbers, the magnetic fluxes $n$ and $N$ and the $q$ charge. The
sigma-model limit, even though it is definitely useful for understanding
certain properties of this soliton, looses some information about it. The
Hopf charge $k$ is equal to the product of the two magnetic fluxes $nN$. In
particular, the sigma-model limit looses the information about the
possibility of higher-genus solitons. For higher genera the Hopf charge is
\begin{equation}
k=\sum_{i=1}^g n_{i}N_{i} \, .
\end{equation}


The analysis of $Q$ tori and higher-genus solitons in the bulk theory is
very complicated. It is not difficult to write the ansatz for the profile
functions and for their boundary conditions but a really hard task is to
solve the equation of motions which cannot be reduced to ordinary
differential equations.

To analyze aspects of this complicated object, we have attacked the problem
in a ``three-fold'' way. The first way is to consider the low-energy
effective action on the worldvolume of the domain wall. The second framework
is based on the bulk theory in the the limit $\Delta m \ll g\sqrt{\xi}$
where the bulk theory reduces to a nonlinear sigma model CP(1) and the $Q$
torus is akin to the Hopf Skyrmion. The third way is to consider the bulk
theory in the thin-edge approximation $\Delta m \gg g\sqrt{\xi}$. In all
these three frameworks we proved the existence  of   $Q$
tori.



An important aspect we should
discuss is that of stability of the $Q$-torus. As we have already noted,
the
issue of the $Q$ charge radiation is essential for understanding whether or not
$Q$-torus is a stable object. In fact, it turns out that the only global
charge we can measure at infinity is the $Q$ charge.  We addressed this
question in Sect.~\ref{qchar}, considering the mass dependence versus
$q$ and  comparing it with the mass of  $q$ elementary
quanta. In a number of soliton problems, such as those of $Q$-kink or  
$Q$-cylinder, we have a powerful argument that answers  the 
question of stability:
a lower bound  for the energy which is a function of the
conserved charges. If the soliton saturates the bound then this is a 
proof of stability. 

Unfortunately, we have no such lower bound for
 $Q$ tori. In this case the issue of the $Q$ charge radiation must be addressed
directly. Analysis of Sect.~\ref{qchar} is a step in this
direction,
 but certainly this is not a complete story. First of all,
because we   consider  only certain limits  in the parameter space in
which we can compute the masses directly. Second, because we should
also consider the issue of stability under all kinds of
fluctuations around the equilibrium position, in particular,
those that break  the cylindrical symmetry. This remains to be a task for the future.

In a sense, the $Q$ tori we study are reminiscent of the ``vorton''
discovered in the context of superconducting cosmic strings
\cite{vorton}. In the latter case we have a string with some internal
degrees of freedom; the vorton is a closed and static
configuration stabilized by some charge and angular momentum of the
flowing degrees of freedom. The question  of vorton stability
is also a delicate issue \cite{stability}.



We want to stress that our soliton, although studied in a specific
environment of ${\mathcal{N}}=2\,$\, SQED can appear also in many other
contexts. The basic ingredients for building of $Q$ tori and higher-genus
solitons of this type are domain walls (or $(2+1)$-dimensional branes) with
a U(1) phase on its worldvolume (or, equivalently, a U(1) gauge field). In
the the field-theoretic context the theory we studied is the \emph{simplest}
one that supports such kind of solitons. This assertion is easy to
understand from the fact that \textit{all} scalar fields in the theory at
hand are excited and play a role in the stability of the $Q$ tori.


The task we had set up is accomplished here. A large number of interesting
problems related to supersymmetric $Q$ solitons remain open for future
research.


\section*{Acknowledgments}


S.B.~was funded by the Marie Curie grant MEXT-CT-2004-013510. S.B. wants to
thank FTPI for the hospitality in the fall of 2006, when part of this work
was done. The work of M.S. was supported in part by DOE grant
DE-FG02-94ER408.


\begin{thebibliography}{99}
\addcontentsline{toc}{section}{References}

\bibitem{Shifman2002} {\small M.~Shifman and A.~Yung,
Phys.\ Rev.\ D \textbf{67}, 125007 (2003) [hep-th/0212293].
}

\bibitem{Sakai2005} {\small N.~Sakai and D.~Tong,
JHEP \textbf{0503}, 019 (2005) [hep-th/0501207].
}

\bibitem{Hanany} {\small A.~Hanany, M.~J.~Strassler and A.~Zaffaroni,
Nucl.\ Phys.\ B \textbf{513}, 87 (1998) [hep-th/9707244].
}

\bibitem{Vainshtein} {\small \ A.~I.~Vainshtein and A.~Yung,
Nucl.\ Phys.\ B \textbf{614}, 3 (2001) [hep-th/0012250].
}

\bibitem{rp} {\small M.~Shifman and A.~Yung, \emph{Supersymmetric Solitons
and How They Help Us Understand non-Abelian Gauge Theories}, hep-th/0703267.
}

\bibitem{GTT} {\small J.~P.~Gauntlett, D.~Tong and P.~K.~Townsend,
Phys.\ Rev.\ D \textbf{64}, 025010 (2001) [hep-th/0012178].
}

\bibitem{Tw} {\small D.~Tong, 
Phys.\ Rev.\ D \textbf{66}, 025013 (2002) [hep-th/0202012].
}

\bibitem{GPTT} {\small J.~P.~Gauntlett, R.~Portugues, D.~Tong and
P.~K.~Townsend, 
Phys.\ Rev.\ D \textbf{63}, 085002 (2001) [hep-th/0008221].
}

\bibitem{arai} {\small M.~Arai, M.~Naganuma, M.~Nitta and N.~Sakai,
Nucl.\ Phys.\ B \textbf{652}, 35 (2003) [hep-th/0211103]; \emph{BPS wall in $%
{\mathcal{N}}=2\,$ SUSY Nonlinear Sigma Model with Eguchi--Hanson Manifold},
in \textsl{Garden of Quanta}, Ed. A. Arafune et. al, (World Scientific,
Singapore, 2003) p. 299 [hep-th/0302028]. }

\bibitem{nsvzr} {\small V.~A.~Novikov, M.~A.~Shifman, A.~I.~Vainshtein and
V.~I.~Zakharov,
Phys.\ Rept.\ \textbf{116}, 103 (1984).  
}

\bibitem{shl} {\small M. Shifman, \emph{Supersymmetric Solitons and Topology,%
} Lect. Notes Phys. \textbf{659}, 237-284, \textsl{2005} (Springer, 2005,
Eds. E. Bick and F. Steffen). }

\bibitem{P77} {\small A.~M.~Polyakov,
Nucl.\ Phys.\ B \textbf{120}, 429 (1977). 
}

\bibitem{ano} {\small A.~Abrikosov, Sov.~Phys. JETP \textbf{32} 1442 (1957)
[Reprinted in \emph{Solitons and Particles}, Eds. C. Rebbi and G. Soliani
(World Scientific, Singapore, 1984), p. 356]; H.~Nielsen and P.~Olesen,
Nucl.~Phys. \textbf{B61} 45 (1973) [Reprinted in \emph{Solitons and Particles%
}, Eds. C. Rebbi and G. Soliani (World Scientific, Singapore, 1984), p.
365]. }

\bibitem{ds} {\small G.~R.~Dvali and M.~A.~Shifman,
Nucl.\ Phys.\ B \textbf{504}, 127 (1997) [hep-th/9611213];
G.~R.~Dvali and M.~A.~Shifman,
in the L. Okun Festschrift, Phys.\ Rept.\ \textbf{320}, 107 (1999)
[hep-th/9904021]. 
}

\bibitem{Coleman:1985ki} {\small S.~R.~Coleman, 
Nucl.\ Phys.\ B \textbf{262}, 263 (1985), (E)\ B \textbf{269}, 744 (1986)]. }

\bibitem{Abraham:1991ki} {\small E.~Abraham,
Phys.\ Lett.\ B \textbf{278}, 291 (1992); } {\small E.~R.~C.~Abraham and
P.~K.~Townsend, 
Phys.\ Lett.\ B \textbf{291}, 85 (1992); } {\small E.~R.~C.~Abraham and
P.~K.~Townsend, 
Phys.\ Lett.\ B \textbf{295}, 225 (1992). }

\bibitem{Leese} {\small \ R.~A.~Leese, 
Nucl.\ Phys.\ B \textbf{366}, 283 (1991). 
}

\bibitem{SVZ} {\small M.~Shifman, A.~Vainshtein and R.~Zwicky,
J.\ Phys.\ A \textbf{39}, 13005 (2006) [hep-th/0602004].}

\bibitem{tongreview} {\small D.~Tong, \emph{TASI Lectures on Solitons,}
hep-th/0509216.} 

\bibitem{Bolognesi:2005} {\small S.~Bolognesi,
Nucl.\ Phys.\ B \textbf{730}, 127 (2005) [hep-th/0507273];
Nucl.\ Phys.\ B \textbf{730}, 150 (2005) [hep-th/0507286].}

\bibitem{Bolognesi:2005rj} {\small S.~Bolognesi and S.~B.~Gudnason,
Nucl.\ Phys.\ B \textbf{741}, 1 (2006) [hep-th/0512132];
Nucl.\ Phys.\ B \textbf{754}, 293 (2006) [hep-th/0606065].}

\bibitem{BP} {\small A.~M.~Polyakov and A.~A.~Belavin,
JETP Lett.\ \textbf{22}, 245 (1975).}

\bibitem{Faddeev:1996zj} {\small L.~D.~Faddeev and A.~J.~Niemi,
Nature \textbf{387}, 58 (1997) [hep-th/9610193]. }

\bibitem{Faddeev:1998eq} {\small L.~D.~Faddeev and A.~J.~Niemi,
Phys.\ Rev.\ Lett.\ \textbf{82} (1999) 1624 [hep-th/9807069]. }

\bibitem{Bolognesi:2007ut} {\small S.~Bolognesi and M.~Shifman,
Phys.\ Rev.\ D \textbf{75}, 065020 (2007) [hep-th/0701065].}

\bibitem{Derrick} {\small G.~H.~Derrick,
J.\ Math.\ Phys.\ \textbf{5}, 1252 (1964).}

\bibitem{joh} {\small C.V. Johnson, \textsl{D Branes}, (Cambridge University
Press, 2003). }

\bibitem{Mateos:2001qs} {\small D.~Mateos and P.~K.~Townsend,
Phys.\ Rev.\ Lett.\ \textbf{87} (2001) 011602 [hep-th/0103030]. }

\bibitem{Lee}  {\small \ S.~Kim, K.~M.~Lee and H.~U.~Yee,  \emph{Supertubes
in field theories,}  hep-th/0603179.  
}

\bibitem{vorton}
R.~L.~Davis and E.~P.~S.~Shellard,
  Phys.\ Lett.\  B {\bf 209}, 485 (1988).

 R.~L.~Davis and E.~P.~S.~Shellard,
  Nucl.\ Phys.\  B {\bf 323}, 209 (1989).

\bibitem{stability}
  Y.~Lemperiere and E.~P.~S.~Shellard,
  Phys.\ Rev.\ Lett.\  {\bf 91}, 141601 (2003)
  [arXiv:hep-ph/0305156].

\end{thebibliography}
\end{document}